\documentclass[fleqn,usenatbib]{mnras}

\usepackage{newtxtext,newtxmath}
\usepackage[T1]{fontenc}

\DeclareRobustCommand{\VAN}[3]{#2}
\let\VANthebibliography\thebibliography
\def\thebibliography{\DeclareRobustCommand{\VAN}[3]{##3}\VANthebibliography}

\usepackage{graphicx}	% Including figure files
\usepackage{amsmath}	% Advanced maths commands

\def\oii{O\,\textsc{ii}}
\def\oiii{O\,\textsc{iii}}
\def\sii{S\,\textsc{ii}}

\def\ciii{C\,\textsc{iii}}
\def\civ{C\,\textsc{iv}}

\def\heii{He\,\textsc{ii}}
\def\nii{N\,\textsc{ii}}
\def\niii{N\,\textsc{iii}}

\def\nev{Ne\,\textsc{v}}

\def\tcra{\textcolor{black}}
\def\tcrb{\textcolor{black}}
\def\tcrc{\textcolor{black}}
\def\tcrd{\textcolor{black}}
\def\tcre{\textcolor{black}}
\def\tcrf{\textcolor{black}}

\usepackage{lineno}
\title[CNO Abundances of \tcra{the JADES} High-z AGN Stacks]{\tcre{JADES: Nitrogen Enhancement in High-Redshift Broad-Line Active Galactic Nuclei
}}

\author[Y. Isobe et al.]{Yuki Isobe,$^{1,2,3}$\thanks{E-mail: yi264@cam.ac.uk (YI)}
Roberto Maiolino,$^{1,2,4}$
Francesco D'Eugenio,$^{1,2}$
Mirko Curti,$^{5}$
Xihan Ji,$^{1,2}$
Ignas Juod\v{z}balis,$^{1,2}$
\newauthor
Jan Scholtz,$^{1,2}$
Anne Feltre,$^{6}$
St\'ephane Charlot,$^{7}$
Hannah \"Ubler,$^{8}$
Andrew J.\ Bunker,$^{9}$
Stefano Carniani,$^{10}$
\newauthor
Emma Curtis-Lake,$^{11}$
Zhiyuan Ji,$^{12}$
Nimisha Kumari,$^{13}$
Pierluigi Rinaldi,$^{12}$
Brant Robertson,$^{14}$
Chris Willott,$^{15}$
\newauthor
and Joris Witstok$^{16,17}$
\\
$^{1}$Kavli Institute for Cosmology, University of Cambridge, Madingley Road, Cambridge, CB3 0HA, UK\\
$^{2}$Cavendish Laboratory, University of Cambridge, 19 JJ Thomson Avenue, Cambridge, CB3 0HE, UK\\
$^{3}$Waseda Research Institute for Science and Engineering, Faculty of Science and Engineering, Waseda University, 3-4-1, Okubo, Shinjuku, Tokyo 169-8555, Japan\\
$^{4}$Department of Physics and Astronomy, University College London, Gower Street, London WC1E 6BT, UK\\
$^{5}$European Southern Observatory, Karl-Schwarzschild-Strasse 2, 85748 Garching, Germany\\
$^{6}$INAF-Osservatorio Astrofisico di Arcetri, Largo E. Fermi 5, 50125 Firenze, Italy\\
$^{7}$Sorbonne Universit\'e, CNRS, UMR 7095, Institut d'Astrophysique de Paris, 98 bis bd Arago, 75014 Paris, France\\
$^{8}$Max-Planck-Institut f\"ur extraterrestrische Physik (MPE), Gie{\ss}enbachstra{\ss}e 1, 85748 Garching, Germany\\
$^{9}$Department of Physics, University of Oxford, Denys Wilkinson Building, Keble Road, Oxford OX1 3RH, UK\\
$^{10}$Scuola Normale Superiore, Piazza dei Cavalieri 7, I-56126 Pisa, Italy\\
$^{11}$Centre for Astrophysics Research, Department of Physics, Astronomy and Mathematics, University of Hertfordshire, Hatfield AL10 9AB, UK\\
$^{12}$Steward Observatory, University of Arizona, 933 North Cherry Avenue, Tucson, AZ 85721, USA\\
$^{13}$AURA for European Space Agency, Space Telescope Science Institute, 3700 San Martin Drive. Baltimore, MD, 21210, USA\\
$^{14}$Department of Astronomy and Astrophysics University of California, Santa Cruz, 1156 High Street, Santa Cruz CA 96054, USA\\
$^{15}$NRC Herzberg, 5071 West Saanich Rd, Victoria, BC V9E 2E7, Canada\\
$^{16}$Cosmic Dawn Center (DAWN), Copenhagen, Denmark\\
$^{17}$Niels Bohr Institute, University of Copenhagen, Jagtvej 128, DK-2200, Copenhagen, Denmark
}

\date{Accepted 2025 May 14. Received 2025 May 09; in original form 2025 February 17}

\pubyear{\the\year{}}

\begin{document}
\label{firstpage}
\pagerange{\pageref{firstpage}--\pageref{lastpage}}
\maketitle

% Abstract of the paper (200 words for Letters)
\begin{abstract}
The unexpectedly high nitrogen-to-oxygen (N/O) ratios observed in high-redshift ($z$) galaxies have challenged our understanding of early star formation. Notably, many of these nitrogen-rich galaxies \tcrb{show} signatures of active galactic nuclei (AGNs), suggesting a possible connection between black hole formation and nitrogen enrichment.
To explore \tcra{this connection}, we analyse stacked spectra of $z=4-7$ \tcra{broad-line and narrow-line} AGNs using deep NIRSpec data from the JADES survey.
We \tcra{identify a significant \niii] quintuplet and a high electron density ($\sim10^{4}$\,cm$^{-3}$) only in the broad-line AGN stack}, indicating nitrogen-rich (\tcrb{$\log(\mathrm{N/C})\simeq0.5$, $\log(\mathrm{N/O})>-0.6$}) and dense gas similar to the high-$z$ nitrogen-rich galaxies.
Our findings suggest that dense nuclear star formation may trap nitrogen-rich gas in proto-globular clusters, in line with the high N/O observed in local globular \tcrb{clusters}; associated runaway stellar collisions could produce intermediate-mass black hole seeds, as predicted by some models and simulations, whose accretion results into AGN signatures. \tcrb{These findings support scenarios connecting the early black hole seeding and  growth to merging processes within and between proto-globular clusters in primeval galaxies.}
\end{abstract}


%%%%%%%%%%%%%%%%%%%%%%%%%%%%%%%%%%%%%%%%%%%%%%%%%%

%%%%%%%%%%%%%%%%% BODY OF PAPER %%%%%%%%%%%%%%%%%%

\section{Introduction} \label{sec:intro}

\tcrb{Chemical abundance ratios of the galaxy interstellar medium, such as nitrogen-to-oxygen (N/O), trace accumulated yields of dying stars with varying masses, serving as key indicators of star formation history.}
Observations \citep[e.g.,][]{Izotov2006} and models \citep{Vincenzo2016} show that
\tcrb{N/O generally increases with gas-phase metallicity (O/H): primary \tcrc{nitrogen (N)} production by massive stars with predominant \tcrc{oxygen (O)} from core-collapse supernovae (CCSNe) during $12+\log(\mathrm{O/H})\lesssim8.2$, secondary N production from asymptotic giant branch (AGB) stars during $12+\log(\mathrm{O/H})\sim8.2-8.6$, and the galactic wind removing O during $12+\log(\mathrm{O/H})\gtrsim8.6$.}
\tcrb{Although \tcrd{the N/O value has a large scatter at $12+\log(\mathrm{O/H})\sim7.6-8.3$ \citep{Kumari2018}} possibly due to Wolf-Rayet (WR) stars \citep[e.g.,][]{Kumari2018} or metal-poor inflow \citep[e.g.,][]{Amorin2010}, their N/O values rarely exceed the solar abundance \citep[cf. Mrk996;][]{Telles2014}.}

However, the \textit{James Webb Space Telescope} (JWST; \citealt{Gardner2023,Rigby2023}) has identified $\sim10$ galaxies with supersolar N/O ratios at \tcra{$z\gtrsim5$} \citep[e.g.,][]{Bunker2023,Ji2024,Schaerer2024}.
These galaxies are also N-rich with respect to \tcrc{carbon (C)}, indicative of CNO-cycle processed gas (\citealt{Isobe2023c}, see also \citealt{Arellano2024}).
Notably, many of these N-rich galaxies have high electron densities $n_{\mathrm{e}}$ \citep[e.g.,][]{Yanagisawa2024,Topping2025}, high stellar mass surface densities \citep[e.g.,][]{Schaerer2024}, and/or compact morphology \citep{Harikane2025}.
From these facts, the possibility that nitrogen is enriched in dense starbursts has been actively discussed \citep[e.g.,][]{Topping2025}.
This might also suggest the link between N-rich galaxies and proto-globular clusters \tcrb{\citep[GCs; e.g.,][]{Senchyna2024,Renzini2023,DAntona2023}}.

It is also worth mentioning that about half of the N-rich galaxies are reported to have signatures of active galactic nuclei (AGNs), such as \tcra{broad-line regions \tcrb{(BLRs)}} for GS\_3073 \citep{Ubler2023,Ji2024}, CEERS\_01019 \citep{Larson2023}, and \tcra{CANUCS-LRD-z8.6 \citep{Tripodi2024}}, ultra-dense gas for GN-z11 (\citealt{Maiolino2024a}\tcrb{; but see also \citealt{Alvarez2025}}),  [\nev]$\lambda$3426 for JADES-GS-z9-0 \citep{Curti2024}, and X-rays in GHZ9 \citep{Napolitano2024b}.
Therefore, there might be a connection between N-enrichment, proto-GC formation, and AGN activity. Indeed, runaway mergers inside star clusters are thought to be a major path to provide massive black hole (BH) seeds \citep[e.g.,][]{Inayoshi2020,Partmann2025,Rantala2024}.

JWST has discovered many \tcrb{AGN candidates} at high $z$ \tcrb{\citep[e.g.,][]{Kocevski2023,Harikane2023,Maiolino2024c,Taylor2024,Mazzolari2024}}, allowing us to verify whether high-$z$ AGNs are N-rich.
Although most of these AGNs are not bright enough to \tcrb{produce observable} N lines, it may be possible to detect them by stacking available spectra.

In this Letter, we report the CNO abundances of high-$z$ AGNs using their stacked spectra observed with the JWST/Near Infrared Spectrograph (NIRSpec; \citealt{Boker2022,Jakobsen2022}).
We explain our data and sample in Section \ref{sec:datsamp}, analysis in Section \ref{sec:ana}, results and discussions in Section \ref{sec:resdis}, and conclusions in \ref{sec:con}.
Throughout this Letter, we \tcra{refer to
\heii$\lambda$1640 as \heii,
\oiii]$\lambda\lambda$1661,1666 as \oiii],
\niii]\tcrb{$\lambda$}1747-1754 as \niii],
\ciii]$\lambda\lambda$1907,1909 as \ciii]\tcrb{, and}
[\oii]$\lambda\lambda$3727,3729 as [\oii]
for simplicity.}
Throughout this Letter, we use the solar abundance ratios of \citet{Asplund2021}.

\section{Data and Sample} \label{sec:datsamp}

We analyse the data obtained by the \tcrb{JWST Advanced Deep Extragalactic Survey (JADES; \citealt{Bunker2020,Rieke2020,Eisenstein2023})}, using the JWST/NIRSpec micro-shutter array (MSA; \citealt{Jakobsen2022}; \citealt{Ferruit2022}).
JADES consists of six programmes (PIDs 1180, 1181, 1210, 1286, 1287, and 3215) in GOODS-S and GOODS-N.
\tcra{In this Letter, we use the JADES data \tcrb{processed} by November 2024, which include complete datasets of the four programmes (PIDs 1180, 1181, 1210, and 3215 \tcrb{-- see \citealt{Bunker2024,DEugenio2025,Eisenstein2023b}}) and part of the two programmes (PIDs 1286 and 1287).}
\tcre{The JADES has NIRSpec data with $R\sim100$ and $R\sim1000$ (R100 and R1000, hereafter) covering $\sim$1--5\,$\mu$m, which we use in this Letter.}
%As the exposure time for the \tcre{R100} data was generally longer than that for the \tcre{R1000} data by a factor of $1-4$ \citep{DEugenio2025}, we derive most of the emission line ratios using the \tcre{R100} data.
%\tcrb{We} use \tcre{R1000} data in the less noisy \tcrb{restframe} optical range to measure [\oiii]$\lambda$4363/[\oiii]$\lambda$5007 and [\sii]$\lambda$6716/[\sii]$\lambda$6731 for better line decompositions.

The observed JADES data were reduced by the NIRSpec GTO Team, using the data reduction pipeline developed by the ESA NIRSpec Science Operations Team \citep{Ferruit2022} and the NIRSpec GTO Team \citep{Oliveira2018}, whose details are presented in \citet{DEugenio2025}.
\tcra{Our 1D spectra are a re-extraction of the JADES data.
Our main targets are high-$z$ AGNs, which should be compact.
To increase the S/N in the blue regions of the spectrum where the JWST point spread function is narrowest, we adopt a 3-pixel box-car aperture (instead of the standard 5-pixel aperture presented in \citealt{DEugenio2025}).}
\tcra{The standard pipeline uses error propagation, with variance-conserving resampling to (conservatively) account for correlated noise \citep{Dorner2016}.}
We also use spectra whose $z$ values are reliably determined \tcra{from} multiple emission lines (i.e., flag values of 6, 7, or 8; \citealt{DEugenio2025}).

We use a sample of \tcrb{broad-line AGNs (aka Type-1 AGNs)}, which are selected by \citet{Juodzbalis2025} using the full JADES NIRSpec data based on a broad-component detection \tcrb{($S/N>5$)} in H$\alpha$ (\tcrb{cf.} \citealt{Juodzbalis2024}). 
These H$\alpha$-selected Type-1 AGNs are located at $z<7$ due to the wavelength coverage of NIRSpec.
%\tcra{We focus on $z>4$, where the \tcre{R100} spectra fully cover key emission lines of \niii], \oiii], and \ciii] with similar \tcrb{resolution}.}
We focus on $z>4$, where the \tcre{R100} spectra fully cover key emission lines of \niii], \oiii], and \ciii].
This Letter targets \tcra{all these Type-1 AGNs at $z=4-7$}, whose number is twenty.

\tcrb{We also construct a ``Type-2 AGN'' sample using Type-2 AGN candidates, which} \tcre{\citet{Scholtz2023} selected} from part of the JADES programmes (PIDs 1210 and 3215) \tcrb{based on} emission line diagnostics of [\nii]$\lambda$6583/H$\alpha$ vs. [\oiii]$\lambda$5007/H$\beta$ \citep[][]{Baldwin1981}\tcrb{,} [\sii]$\lambda$6731/H$\alpha$ vs. [\oiii]$\lambda$5007/H$\beta$ \tcra{\citep{VO87}}\tcrb{, and} those based on high-ionisation lines \tcra{(e.g., \heii; \citealt{Hirschmann2023})}.
We use \tcra{all these Type-2 \tcrb{AGN candidates} at $z=4-7$}, whose number is eighteen.
\tcrb{Note that there is only one overlap between our Type-1 and Type-2 AGN samples, which means that most of our Type-2 AGNs are narrow-line AGNs.}

Additionally, \tcra{excluding \tcrb{our} Type-1 or Type-2 AGNs, we obtain 665 galaxies at $z=4-7$ where no evidence of AGN has been identified.}
Hereafter, we refer to this galaxy sample as the ``\tcra{Non-AGN}'' sample.
\tcrb{Note that our three samples have similar median $z$ values of 5.2--5.3.}

\section{Analysis} \label{sec:ana}
\subsection{Emission Line Fitting for Individual Galaxies} \label{subsec:indiv}
%We conduct emission line fitting for individual galaxies to obtain redshifts fully based on the \tcre{R100} spectra.
We conduct emission line fitting for individual galaxies to obtain redshifts \tcre{based on the \tcre{R100} and \tcre{R1000} spectra independently}.
We model the spectra around H$\beta$ and [\oiii]$\lambda\lambda$4959,5007 with 3 Gaussian functions and \tcra{a constant continuum level}.
We use the $z$ values with visual inspection \citep{DEugenio2025} as an initial guess.
We fix the ratio of \tcrb{[\oiii]$\lambda$5007/[\oiii]$\lambda$4959 to 2.98 \citep{Kojima2020}}.
We also use the same width for H$\beta$ and [\oiii]$\lambda\lambda$4959,5007.
\tcrb{There are 64 overlaps \tcre{(2 of Type-1, 12 of Type-2, and 50 of Non-AGN sources)} between our sample and \citet{Bunker2024}, whose $z$ values based on the \tcre{R100} data are different by only at most 0.1\%.}
This analysis provides not only the $z$ values but also [\oiii]$\lambda$5007 fluxes ($F$([\oiii])), which are used to normalise the spectra in Section \ref{subsec:stk}.

\subsection{Stacking} \label{subsec:stk}
\begin{figure*}
	%\includegraphics[width=\columnwidth]{example}
	\includegraphics[width=0.9\textwidth]{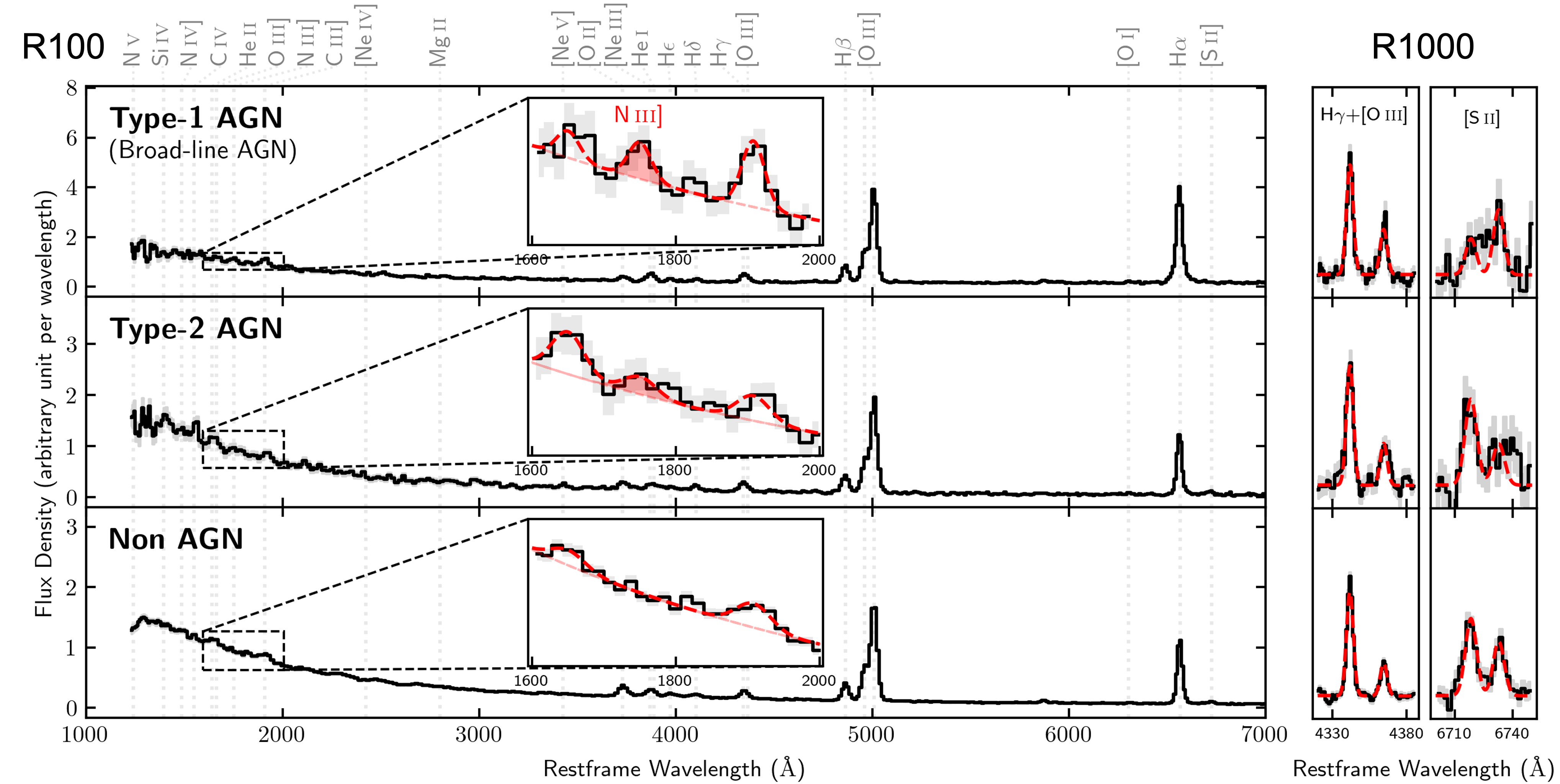}
    \caption{\tcrf{\textbf{Left:} R100} \tcre{composite spectra of the median stacking} (black solid line) \tcrb{with the error (gray shade)} of the Type-1 AGN (top), the Type-2 AGN (middle), and the \tcra{Non-AGN} samples (bottom).
    The inset panels show fitting results of the UV emission lines (red dashed curve), highlighting \niii] lines with the red shade. \tcrf{\textbf{Right:} R1000 stacked spectra around H$\gamma$+[\oiii]$\lambda$4363 and [\sii]. The symbols are the same as the left figure.}}
    \label{fig:stkspec}
\end{figure*}

We \tcra{produce} stacked spectra \tcre{of the \tcre{R100} and \tcre{R1000} data in the same manner}.
We shift the spectra to the restframe and then resample them to a common wevelength grid, for which the spectral pixel size is set to half of \tcrb{the full-width half maximum based on} the NIRSpec spectral resolution.
We use the spectral resolution in the case of compact/point-like sources (hence smaller than the shutter width; \citealt{deGraaff2024}), which is generally two times better than the nominal resolution, which \tcrb{applies only to} the case of uniformly illuminated shutter.
The resampling to the common wavelength grid is done using \texttt{spectres} \citep{Carnall2017}, which also propagates the errors of the individual spectra to those of the resampled spectra.

Finally, we renormalise the resampled spectra by the $F$([\oiii]) values obtained in Section \ref{subsec:indiv}, and then \tcrb{we define the stacked spectrum by taking the median of the normalised spectra} \tcre{without any weighting} at each wavelength of the common $\lambda_{\mathrm{rest}}$ grid\tcre{, which underlies all the results reported in this Letter}.
%This normalisation and the use of median values help mitigate the influence of a few bright outliers, ensuring that the resulting spectra more accurately represent the average properties of the individual galaxies.
\tcre{This stacking method helps mitigate the influence of a few bright outliers, ensuring that the resulting spectra more accurately represent the average properties of the individual galaxies.}
We obtain errors of the stacked spectrum by performing a Monte Carlo simulation.
We create 1000 stacked spectra based on the individual spectra randomly fluctuated by their errors under the assumption of the normal distribution, and calculate the standard deviation at each wavelength.
\tcrb{When} we adopt this method to the mean stacking, it can reproduce the errors propagated from the individual error spectra as done by \citet{Steidel2016}.
\tcre{Bootstrap resampling is another method to account for the errors of the stacked spectra.
However, we do not use bootstrap errors for emission lines, as they overestimate flux variance by including the strong effect of variation of the underlying continuum emission from source to source.}
The stacked spectra are \tcrb{shown} in Fig.~\ref{fig:stkspec}.

\subsection{Emission Line Measurement} \label{subsec:emis}
We conduct emission line fitting for the stacked spectra in the following three wavelength regions (\tcrb{UV: }1600--2000, \tcrb{blue: }3600--5200, \tcrb{red: }6200--7300 \AA).
%We fit the Gaussian function to each emission line (\heii+\oiii], \niii], and \ciii] for the \tcrb{UV}, [\oii], [\neiii]$\lambda$3869, \hei$\lambda$3889, [\neiii]$\lambda$3967+\tcra{H$\zeta$}, H$\delta$, H$\gamma$, [\oiii]$\lambda$4363, H$\beta$, [\oiii]$\lambda\lambda$4959,5007 for the \tcrb{blue}, [\oi]$\lambda$6300, H$\alpha$+[\nii]$\lambda\lambda$6548,6583, [\sii]$\lambda\lambda$6716,6731, and [\ariii]$\lambda$7136 for the \tcrb{red} wavelength region).
\tcre{As the exposure time for the \tcre{R100} data was generally longer than that for the \tcre{R1000} data by a factor of $1-4$ \citep{DEugenio2025}, we derive key emission line fluxes (\heii+\oiii], \niii], \ciii], [\oii], H$\beta$, [\oiii]$\lambda\lambda$4959,5007, H$\alpha$+[\nii]$\lambda\lambda$6548,6583) using the \tcre{R100} data, except for [\oiii]$\lambda$4363/[\oiii]$\lambda$5007 and [\sii]$\lambda$6716/[\sii]$\lambda$6731.
We obtain these ratios by resolving  [\oiii]$\lambda$4363 from H$\gamma$ and [\sii]$\lambda$6716 from [\sii]$\lambda$6731 with the \tcre{R1000} data in the less noisy restframe optical range.
All our Type-1 AGNs have \tcre{R1000} data, only one of which does not cover [\oiii]$\lambda$4363, while the others cover both [\oiii]$\lambda$4363 and [\sii].
Also, 95\% of our Type-2 and Non AGNs have \tcre{R1000} data, 80--90\% of which cover [\oiii]$\lambda$4363 and [\sii].}
Due to the difficulty of decomposing \oiii] from \heii\ with the \tcre{R100} spectra, we treat \heii+\oiii] as \oiii], which serves as an upper limit \tcrb{of the \oiii] flux}.
We also add a broad Gaussian component for the H$\alpha$ of the Type-1 AGN stack at the same wavelength as the narrow component.
\tcre{At $z=4-7$ with \tcre{R100} data, the ratio of maximum to minimum resolution is only $\sim1.3$ in the rest-frame UV but 2.2–2.4 in the rest-frame optical. However, we find that the cumulative integral of H$\alpha$ in the Non-AGN stack, which excludes Type-1 AGNs, exceeds that from Gaussian fitting by only $\sim$3\%.
This suggests that profile smearing due to resolution differences has a negligible impact on our line flux measurement.}
We use the same width for the emission lines\tcrb{, and} model the continuum with \tcrb{a} power law function at each wavelength region.
%\tcre{The best-fit widths of the \tcre{R100} data lie between those from the resolutions for compact \citep{deGraaff2024} and nominal extended sources.}
\tcre{The best-fit widths of the R100 data lie between those of the resolutions for compact/point-like sources and for nominal extended sources.}
We set a detection criterion of $S/N>2$, and calculate $2\sigma$ upper limits for \tcrb{undetected} lines.
The inset panels of \tcrb{Fig.~}\ref{fig:stkspec} highlight the fitting results of the \tcrb{UV} wavelength region.
\tcre{We find a detection of 2.6$\sigma$ (i.e., $P\,\mathrm{value}<0.005$) in the Type-1 AGN stack.
Note that, since the redshift is fixed, we use the line $S/N$ to infer the significance, with no need to account for the look-elsewhere effect.
In contrast, \niii]} is \tcrb{not significantly detected} in the Type-2 AGN stack \tcrb{(1.6$\sigma$)} \tcrb{or} the \tcra{Non-AGN} stack \tcrb{(0.0$\sigma$)}.
\tcrb{We have checked that this finding holds true regardless of the stacking method (median vs. mean) or normalisation parameter (i.e., $F$([\oiii]), \tcre{[\oiii] luminosity,} vs. continuum at restframe 5500--6000 \AA).}
%\tcre{Note that \niii] lines with $S/N=2.4$-2.6, 2.3-3.0, and $>$2 are seen in 3 of 20 Type-1, 2 of 18 Type-2, and $\sim30$ of 665 Non-AGN sources, respectively, using the \tcre{R100} spectra.
%However, even removing these sources, we see a $\sim2\sigma$ \niii] in the Type-1 AGN stack, suggestive of the overall nature.
%Also, the absence of \niii] in our Type-2 and Non-AGN stacks reflects general trends, but does not imply that none of our Type-2 or Non-AGN sources exhibit \niii].
%We will explore properties responsible for \niii] in a separate paper.}
\tcre{
Note that \niii] lines with $S/N=2.4$-2.6 are seen in 3 out of 20 individual R100 spectra of our Type-1 AGNs.
However, even removing these sources, we see a $\sim2\sigma$ \niii] in the Type-1 AGN stack, suggestive of the overall nature.
Similarly, using the R100 data, we find that 2 out of 18 Type-2 AGNs and $\sim30$ out of 665 Non-AGN sources have \niii] lines with $S/N>2$.
We stress that the absence of \niii] in our Type-2 and Non-AGN stacks reflects general trends, but does not imply that none of our Type-2 or Non-AGN sources exhibit \niii].
}
\tcrb{The observed CNO line ratios of the Type-1 AGN stack (\niii]/\ciii$]=0.56\pm0.24$ and \niii]/\oiii$]>0.99$) are comparable to those of the $z\gtrsim5$ N-rich galaxies \citep[e.g.,][]{Bunker2023} and $z\sim2-3$ N-loud quasars \citep{Batra2014}.
\tcre{Notably, our Type-1 AGN stack does not have \civ, in line with most of the Type-1 AGNs found with the JWST, which may be due to the optically-thick disc softening the incident spectrum \citep{Lambrides2024}, possibly associated with high accretion rate.}
Line fluxes are listed in Tab.~\ref{tab:emis}.
}

\begin{table*}
	\centering
	\caption{\tcrb{Emission line fluxes before dust correction based on the \tcre{R100} spectra normalised by the H$\beta$ flux, except for $^{\dagger}$ from the \tcre{R1000} spectra. The upper limits are 2$\sigma$. $^{\ddagger}$: Narrow component.}}
	\label{tab:emis}
    \resizebox{\textwidth}{!}{
	\begin{tabular}{cccccccccc}
		\hline
        Sample & \oiii] & \niii] & \ciii] & [\oii] & H$\beta$ & [\oiii]$\lambda$5007 & H$\alpha$+[\nii]$\lambda\lambda$6548,6583 & [\oiii]$\lambda$4363/[\oiii]$\lambda$5007$^{\dagger}$ & [\sii]$\lambda$6716/[\sii]$\lambda$6731$^{\dagger}$ \\
        \hline
        Type-1 AGN & $<0.25$ & $0.25\pm0.09$ & $0.44\pm0.09$ & $0.29\pm0.03$ & $1.00\pm0.03$ & $5.44\pm0.05$ & $4.41\pm0.06^{\ddagger}$ & $0.030\pm0.003$ & $0.55\pm0.20$\\
        Type-2 AGN & $0.89\pm0.27$ & $<0.47$ & $0.53\pm0.21$ & $0.40\pm0.06$ & $1.00\pm0.04$ & $5.56\pm0.03$ & $2.65\pm0.04$ & $0.032\pm0.004$ & $2.09\pm0.74$\\
        Non AGN & $0.35\pm0.09$ & $<0.17$ & $0.55\pm0.07$ & $0.64\pm0.02$ & $1.00\pm0.01$ & $5.82\pm0.02$ & $3.20\pm0.02$ & $0.024\pm0.001$ & $1.43\pm0.18$\\
        \hline
	\end{tabular}
    }
\end{table*}

\subsection{Nebular \tcra{Properties}} \label{subsec:neb}
\begin{table*}
	\centering
	\caption{Nebular properties. \tcrb{$^{\dagger}$: Showing a 1$\sigma$ upper limit because the \tcre{measured} value reaches the low density limit. The other limits are 2$\sigma$.}}
	\label{tab:neb}
    \resizebox{\textwidth}{!}{
	\begin{tabular}{cccccccccccc}
		\hline
        Sample & $E(B-V)$ & $T_{\mathrm{e}}$\,($10^{4}$\,K) & $\log(n_{\mathrm{e}}/\mathrm{cm^{-3}})$ & $\log(U)$ & $12+\log(\mathrm{O/H})$ & $\log(\mathrm{N/C})$ & \tcrb{ICF($\mathrm{N^{2+}/C^{2+}}$)} & $\log(\mathrm{N/O})$ & \tcrb{ICF($\mathrm{N^{2+}/O^{2+}}$)} & $\log(\mathrm{C/O})$ & \tcrb{ICF($\mathrm{C^{2+}/O^{2+}}$)} \\
        \hline
        Type-1 AGN & $0.43^{+0.07}_{-0.07}$ & $2.19^{+0.14}_{-0.14}$ & $3.94^{+0.76}_{-0.52}$ & $-2.23^{+0.05}_{-0.03}$ & $7.46^{+0.16}_{-0.06}$ & $0.50^{+0.17}_{-0.25}$ & \tcrb{$1.141^{+0.014}_{-0.002}$} & $>-0.58$ & \tcrb{$1.111^{+0.022}_{-0.017}$} & $>-1.08$ & \tcrb{$0.973^{+0.012}_{-0.016}$} \\
        Type-2 AGN & $0.00^{+0.08}_{-0.08}$ & $1.94^{+0.14}_{-0.13}$ & \tcrb{$<1.98^{\dagger}$} & $-2.13^{+0.09}_{-0.04}$ & $7.55^{+0.07}_{-0.06}$ & $<0.57$ & \tcrb{$1.154^{+0.015}_{-0.004}$} & $<-0.71$ & \tcrb{$1.167^{+0.039}_{-0.026}$} & $-1.28^{+0.20}_{-0.27}$ & \tcrb{$1.011^{+0.022}_{-0.022}$} \\
        \tcra{Non AGN} & $0.13^{+0.03}_{-0.03}$ & $1.74^{+0.05}_{-0.05}$ & \tcrb{$<2.41^{\dagger}$} & $-2.33^{+0.02}_{-0.02}$ & $7.68^{+0.03}_{-0.03}$ & $<0.14$ & \tcrb{$1.064^{+0.001}_{-0.002}$} & $<-0.84$ & \tcrb{$1.056^{+0.004}_{-0.004}$} & $-0.99^{+0.13}_{-0.12}$ & \tcrb{$0.992^{+0.005}_{-0.005}$} \\
        \hline
	\end{tabular}
    }
\end{table*}

\begin{figure*}
	%\includegraphics[width=\columnwidth]{example}
    \includegraphics[width=14.5cm]{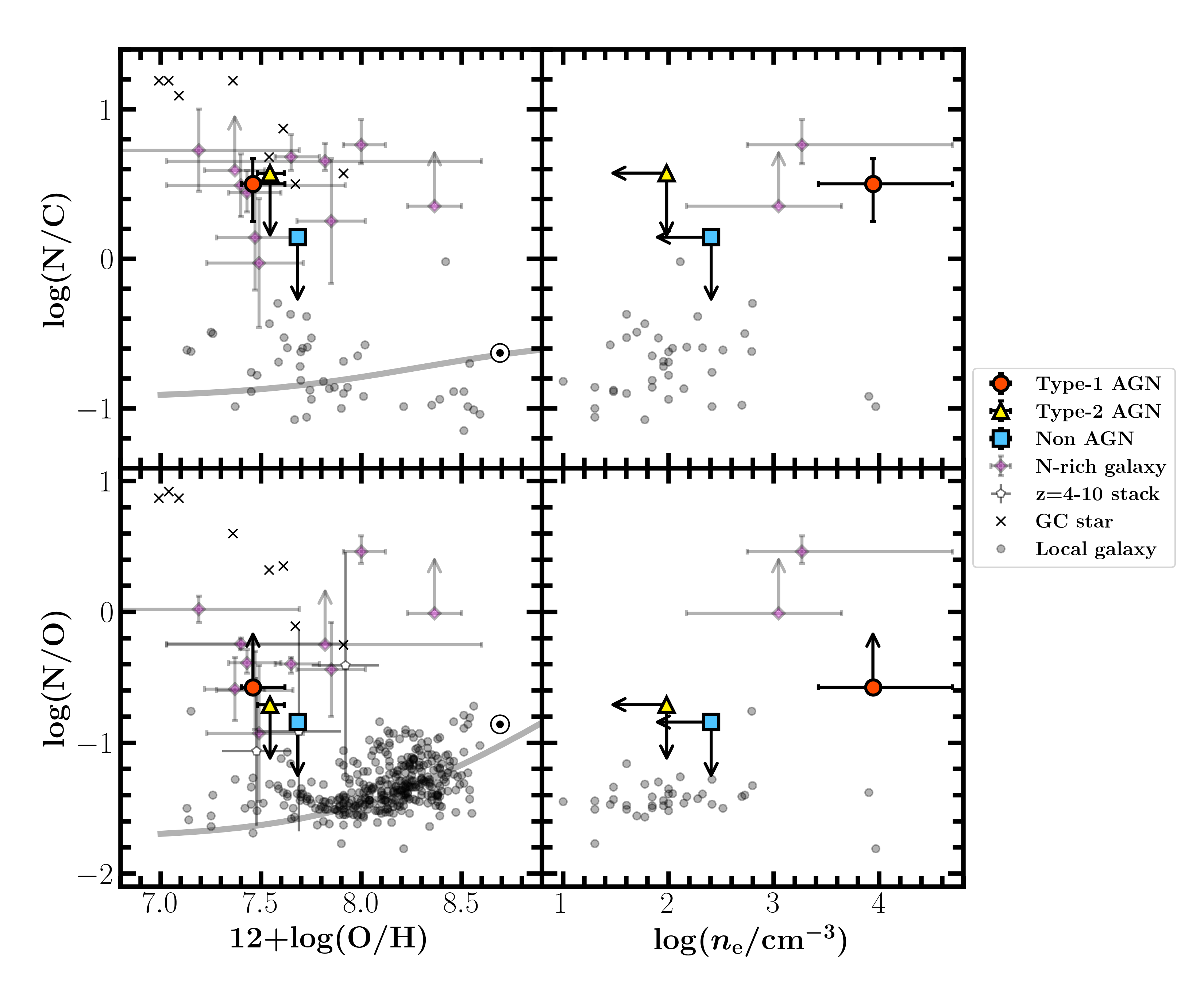}
    \caption{N/C \tcrb{(top)} and N/O ratios \tcrb{(bottom)} as a function of metallicity, \tcra{expressed in terms of oxygen abundance O/H} \tcrb{(left), and $n_{\mathrm{e}}$ of the singly-ionised region (right)}. The red circle corresponds to the Type-1 AGN stack, the yellow triangle to the Type-2 AGN stack, the blue square to the \tcra{Non-AGN} stack, the magenta diamonds to the \tcra{$z\gtrsim5$} N-rich galaxies (\citealt{Cameron2023,Isobe2023c,Topping2024a,Topping2025,Castellano2024,Ji2024,Schaerer2024,Navarro-Carrera2024,Curti2024,Napolitano2024b}; \tcrb{\citealt{Ubler2023} and \citealt{Isobe2023b} for $n_{\mathrm{e}}$}; see also \citealt{Senchyna2024,Marques-Chaves2024,Larson2023}), the white pentagons to the stack of $z=4-10$ galaxies \citep{Hayes2025}, the black crosses to \tcrb{GC} stars \citep{Carretta2005}, the gray dots to local star-forming galaxies \citep{Izotov2006,Berg2016,Berg2019}, the gray curve to the empirical relation based on observations of galaxies and Galactic stars \citep{Nicholls2017}\tcra{, and the circled dot to the solar abundances \citep{Asplund2021}}.}
    \label{fig:abunmet}
\end{figure*}

We derive the colour excess $E(B-V)$, the electron temperature $T_{\mathrm{e}}$(\oiii), and the electron density $n_{\mathrm{e}}$(\sii) as follows.
The $E(B-V)$ values are derived from the H$\beta$/H$\alpha$ ratio, whose intrinsic value under the assumptions of the case B recombination is derived with PyNeb \citep{Luridiana2015}.
We use the same atomic data as \citet{Isobe2023c} except that we use \citet{Lennon1994}'s data for the O$^{2+}$ collisional strength.
We assume the average SMC extinction curve \citep{Gordon2003} with $R_{V}=3.1$, which is consistent with observations \tcrb{of} galaxies at $z\sim1-3$ \citep{Reddy2023a}.
For the Type-1 AGN stack, we use the H$\alpha$ flux value of the narrow component.
The obtained $E(B-V)$ values are consistent with those derived from H$\gamma$/H$\beta$ ratios within 1$\sigma$, where H$\gamma$ is decomposed from [\oiii]$\lambda$4363 using the \tcre{R1000} spectra.
\tcre{The R1000 spectra also indicate that contamination from [\nii]$\lambda$6583 to H$\alpha$ would be at most $\sim3$\% for all samples \citep[cf.][]{Cameron2023b,Sandles2024}, which can only reduce the $E(B-V)$ value by $\lesssim0.03$ and the N/C value by $\lesssim0.02$ dex, and rather increase the N/O value by $\lesssim0.01$ dex.}
We also use PyNeb to derive the $T_{\mathrm{e}}$(\oiii) value from the [\oiii]$\lambda$4363/[\oiii]$\lambda$5007 ratio and the $n_{\mathrm{e}}$(\sii) value from the [\sii]$\lambda$6716/[\sii]$\lambda$6731 ratio.
As the $E(B-V)$, $T_{\mathrm{e}}$(\oiii), and $n_{\mathrm{e}}$(\sii) values slightly depend on each other, we calculate $E(B-V)$, $T_{\mathrm{e}}$(\oiii), and $n_{\mathrm{e}}$(\sii) values iteratively to find their values consistent with each other \citep{Isobe2022}.
We find these values in reasonable ranges of $E(B-V)>0$ \tcrb{mag} and $n_{\mathrm{e}}$(\sii$)=4-50000$\,cm$^{-3}$, the last of which is determined where the [\sii] doublet is sensitive to $n_{\mathrm{e}}$.
We calculate errors of these properties from the 1$\sigma$ errors of the used emission line ratios.

Using the \tcra{final} value of $E(B-V)$, we obtain dust-corrected fluxes of the stacked spectra.
\tcrb{We have checked that the \niii] is detected only in the Type-1 AGN stack even when we correct for the dust reddening based on individual $E(B-V)$ values before the stacking.}

We derive ion abundance ratios of O$^{+}$/H$^{+}$ from [\oii]/H$\beta$, O$^{2+}$/H$^{+}$ from [\oiii]$\lambda$5007/H$\beta$, N$^{2+}$/C$^{2+}$ from \niii]/\ciii], N$^{2+}$/O$^{2+}$ from \niii]/\oiii], and C$^{2+}$/O$^{2+}$ from \ciii]/\oiii]\tcrb{, using $T_{\mathrm{e}}$(\oiii) and $n_{\mathrm{e}}$(\sii) for all ion abundance ratios but $T_{\mathrm{e}}$(\oii) for O$^{+}$/H$^{+}$, where $T_{\mathrm{e}}$(\oii) is derived from $T_{\mathrm{e}}$(\oiii) based on the empirical relation \citep{Garnett1992}.}
\tcra{Similar wavelengths of \niii], \ciii], and \oiii] make their line ratios less sensitive to flux calibration or dust correction\tcrb{, which are conversely critical for e.g., \niii]/[\oiii]$\lambda$5007}.}
We assume $12+\log(\mathrm{O/H})$ to be $12+\log(\mathrm{(O^{+}+O^{2+})/H^{+}})$ \citep[e.g.,][]{Izotov2006}.

To derive element abundance ratios, we calculate ionisation correction factors (ICFs), which are defined as e.g., $\mathrm{N/C}=\mathrm{N^{2+}/C^{2+}\times ICF(N^{2+}/C^{2+})}$.
To obtain $\mathrm{ICF(N^{2+}/C^{2+})}$, $\mathrm{ICF(N^{2+}/O^{2+})}$, and $\mathrm{ICF(C^{2+}/O^{2+})}$, we construct photoionisation models \tcra{based on stellar and AGN \tcrb{ionising spectra}} using Cloudy \citep{Ferland2013} and following the procedure of \citet{Isobe2023c}.
\tcra{For both photoionisation models, we vary O/H within $-2\leq[$O/H$]\leq1$ in 0.25\tcrb{-dex} increments and $U$ within $-3.5\leq\log(U)\leq-0.5$ in 0.25\tcrb{-dex} increments.
We also assume the hydrogen density $n_{\rm H}$ to be 300 cm$^{-3}$ and He/H and metal-to-oxygen ratios to be the solar abundances.
Note that the derived densities are all significantly below the critical densities of lines used for the CNO abundance measurements, which makes the choice of density less important.}

To construct stellar photoionisation models, we use BPASS \citep{Stanway2018} binary stellar \tcrb{spectra} under the assumption of \tcrb{an} instantaneous star-formation history with the stellar age of 10 Myr and an upper star mass cut of 100 $M_{\odot}$ with the \citet{Salpeter1955} initial mass function (IMF).
\tcrb{Different stellar age (1--100 Myr) and upper star mass cut (100 and 300 $M_{\odot}$) can change the ICFs by only $\lesssim0.01$ dex.}
We also construct AGN photoionisation models assuming the AGN radiation model of Cloudy, whose parameter sets are the big-bump temperature of $T_{\rm BB}=1.5\times10^{5}$ K, the X-ray to UV ratio of $\alpha_{\rm OX}=-1.4$, the low-energy slope of the big-bump component of $\alpha_{\rm UV}=-0.5$, and the X-ray component slope of $\alpha_{\rm X}=-1.0$.

We choose the model whose metallicity is the closest to the observed value.
We make the ICF as a function of [\oiii]$\lambda$5007/[\oii] by performing a linear interpolation of the ICF values based on the model [\oiii]$\lambda$5007/[\oii] values.
We then obtain the ICF value at the observed [\oiii]$\lambda$5007/[\oii] value.
\tcra{The obtained ICFs are $\simeq1$ \tcrb{(Tab.~\ref{tab:neb}), indicating only a small ($<0.08$ dex) correction}.
\tcrb{The ICFs can also} change by only $\lesssim0.1$ dex when we add dust that causes the observed extinction of $E(B-V)\lesssim0.4$ \tcrb{mag} only within the modelled ionised region\tcrb{, which is an extreme case because the dust optical depth within the ionised gas is usually very low \citep{Bottorff1998}}.}
We also obtain $\log(U)$ values from [\oiii]$\lambda$5007/[\oii] ratios \tcrb{using the Cloudy models}.

We finally obtain element abundance ratios of N/C, N/O, and C/O from the ion abundance ratios and the ICFs.
We adopt the ICFs based on the AGN photoionisation models for the Type-1 and Type-2 AGN stacks, and those based on the stellar photoionisation models for the \tcra{Non-AGN} stack.
It is worth noting that the derived ICFs are $\sim1$ because the used ions have similar ionisation energies.
This also leads to comparable values of the ICFs based on the AGN and stellar photoionisation models.
We obtain errors of the abundance ratios with a Monte Carlo simulation by repeating the calculation 1000 times based on the flux values \tcrb{randomly perturbed} by their errors.
The nebular properties are listed in \tcrb{Tab.~\ref{tab:neb}}.
\tcrb{Note that the Type-1 AGN stack has a supersolar N/O ratio, which holds true regardless of the choice of dust attenuation laws (\citealt{Gordon2003} vs. \citealt{Calzetti2000}) or line ratios (\niii]/\oiii] or \niii]/[\oiii]$\lambda$5007).}

\section{Result and Discussion} \label{sec:resdis}
\tcrb{The left panels of Fig.~\ref{fig:abunmet} show} CNO abundances of our stacks and N-rich galaxies at \tcra{$z\gtrsim5$}, whose N abundances are similarly derived from restframe-UV N lines \tcrb{(see also [\nii]-based reports at $z\sim4-6$; \citealt{Arellano2024,Stiavelli2025,Zhang2025})}.
The Type-1 AGN stack shows high N/C and N/O ratios comparable to those of the N-rich galaxies \tcra{\citep[e.g.,][]{Cameron2023}} and \tcrb{GC} stars \tcra{\citep{Carretta2005}}.
On the other hand, \tcra{the N/C and N/O ratios of the \tcra{Non-AGN} stack} are lower than many of the N-rich galaxies.

Interestingly, \tcrb{the right panels of Fig.~\ref{fig:abunmet} show} that the Type-1 AGN stack with the high N/C and N/O ratios has $n_{\mathrm{e}}\sim10^{4}$\,cm$^{-3}$.
This value is significantly higher than that of the \tcra{Non-AGN} stack and previous $n_{\mathrm{e}}$ measurements at similar redshifts \citep[e.g.,][]{Isobe2023b,Reddy2023b,Li2025}, and comparable to that of the N-rich galaxies \citep{Ubler2023,Isobe2023b,Larson2023} and simulated young massive \tcrb{star} clusters, promising progenitors of \tcrb{GC}s ($10^{3}$--$10^{5}$\,cm$^{-3}$; \citealt{Inoguchi2024}).

\tcra{
The Type-2 AGN stack has relatively unconstrained N/C and N/O values, which are not conclusive enough to discuss its N enhancement.
However, \tcrb{it is worth mentioning that} its \niii] line is less significant than that of the Type-1 AGN stack and that the $n_{\mathrm{e}}$ value is similarly low to that of the \tcra{Non-AGN} stack.
}

It is worth comparing our results with the stack of $z\sim4-10$ galaxies \citep{Hayes2025}, who also explored N/O in stacked spectra.
We avoid strong conclusions since the uncertainties of their results are large, but we can say that their \tcre{measured} values are closer to the upper limit of our \tcra{Non-AGN} stack than to the lower limit of our Type-1 AGN stack.
This is consistent with the report that \citet{Hayes2025}’s stacked spectra do not show AGN signatures.

\tcrb{What is the origin of the N enhancement in the Type-1 AGN stack?
Various contributors have been proposed to explain early N enhancement, including massive stars ejecting CNO-processed gas via stellar winds \citep[e.g.,][]{Watanabe2024} or tidal disruption events (TDEs;  \citealt{Cameron2023}), and/or chemically differential winds \citep[e.g.,][]{Rizzuti2024}, which may be linked to dense, clustered star formation or supermassive BHs (SMBHs) as in the following evolutionary scenario.}
\underline{Step 1)}: Dense gas in the nuclear region of a galaxy could form stars efficiently, which could lead to \tcrb{several} star clusters (aka proto-\tcrb{GC}s) hosting massive stars \tcrb{or even} supermassive stars \tcrb{(SMSs; $\sim10^{3}-10^{5}\,M_{\odot}$)} through runaway stellar collisions and/or strong gas accretion \citep[e.g.,][]{Gieles2018}.
\tcrb{N-rich gas can be ejected via stellar winds of massive stars, including SMSs \citep[e.g.,][]{Nagele2023,Charbonnel2023,Nandal2024a}, very massive stars \citep[$\sim10^{2}-10^{3}\,M_{\odot}$; ][]{Vink2023}, and WR stars \citep[$\sim25-120\,M_{\odot}$; e.g.,][]{Watanabe2024,Kobayashi2024,Fukushima2024}.}
\tcra{\tcrb{TDEs are also} a natural byproduct of high star densities necessary for runaway collisions.
Gas accretion can spin up the star, \tcrb{favouring the formation of the WR phase} \citep{DallAmico2025}.}
This N-rich gas may be trapped in \tcrb{subsequent generations of} stars in the star cluster, which could link to N overabundance of present-day \tcrb{GC} stars \citep[e.g.,][]{Charbonnel2023,Pascale2023}.
The connection between N enrichment and \tcrb{GC}s will also be discussed in detail by \tcrc{Ji et al. (in prep.)}.
Until Step 1), our scenario is similar to that \tcrb{proposed by} \citet{Topping2025}.
We also note that \tcrb{O can easily be expelled from the dense compact region via SN-driven chemically differential winds, which can result in} enhancing N/O \citep{Vincenzo2016,DAntona2023,Rizzuti2024}.
\underline{Step 2)}: If the \tcrb{stars are} massive enough and/or gas accretion is sufficient, the star cluster would \tcrb{develop} intermediate-mass BHs (IMBHs) with $\sim10^{3}$--$10^{4}\,M_{\odot}$ \tcra{\citep[e.g.,][]{Antti2024,Partmann2025}}.
\tcrb{Indeed, observational signatures of IMBHs have also been found in some GCs \citep[e.g.,][]{Haberle2024}.}
IMBHs can also trigger \tcrb{TDEs} \citep[e.g.,][]{Sakurai2019}.
\underline{Step 3)}: Continuous gas accretion onto the IMBH would create \tcrb{a SMBH} with a dense accretion \tcra{disc}, providing AGN signatures such as \tcrb{BLRs}.
The dense accretion \tcra{disc} could also form \tcrb{VMSs} \citep{Cantiello2021}, which could contribute to N enrichment. \tcrb{Overall, our observational findings support scenarios in which black hole \tcrc{seeding}, and their early growth, is associated with the merging processes within and between proto-GCs in the central regions of primeval galaxies \citep[e.g.][]{Rantala2024,Partmann2025}}.

\tcra{Such extreme N \tcrb{enhancement} would \tcrb{primarily} take place in the central, dense region\tcrb{, possibly the BLR \citep[cf.][]{Maiolino2024a},} but not in the bulk of the host galaxy \citep[see][]{Ji2024}.}
If such objects are observed from the face-on direction \tcrb{without much obscuration by the dusty torus}, N-rich Type-1 AGNs \tcrb{are expected to be seen}.
\tcra{Conversely, from the edge-on direction,
\tcrb{the dusty torus might hide the N-rich central dense region.}
This could be the reason why the Type-2 AGN stack does not have a significant \niii] detection or a high $n_{\mathrm{e}}$ value.}
\tcrb{Considering also that the Type-2 AGN stack shows $E(B-V)=0$ mag, the dust may be strongly localised at the central region.
The possibility that some of our Type-2 AGNs might not be AGNs \citep[see][]{Scholtz2023} would just increase the contribution from the host galaxies, which still fit within our scenario.}

%\tcrb{Given that the \niii] line of the Type-1 AGN stack may be emitted from the BLR, it is worth checking the CNO abundance measurements.
\tcre{Here, we check the CNO abundance measurements of the Type-1 AGN stack, assuming its \niii] is emitted from the BLR.
The reddening of the BLR is generally greater than that of the narrow-line region (NLR) by $\Delta E(B-V)\lesssim0.5$ \citep{Heard2016}, which can make the true N/C and N/O values of the Type-1 AGN stack even higher than the observed values by $\lesssim2$ dex.
Although many parameters can affect chemical abundance measurements in the BLR \citep{Temple2021}, we have checked that the BLR photoionisation models of \citet{Ji2024} can slightly reduce the inferred N/C value by $\sim0.2$ dex and increase the inferred N/O value by $\sim0.3$ dex from the NLR models.
Neither case is likely to change our conclusions.}

Note that our scenario does not exclude the presence of N-rich galaxies without \tcrb{any} AGN signatures.
If the star in Step 1) is not massive enough, gas accretion is not sufficient, or the galaxy is too young to reach Step 3), N-rich galaxies would not \tcrb{develop an} AGN. Additionally, high-z AGN may have a short duty cycle, hence they are observable as active only for relatively short periods \citep{Trinca2024}.
This can explain some of the high-$z$ N-rich galaxies without AGN signatures such as GLASS\_150008 \citep{Isobe2023c} or A1703-zd6 \citep{Topping2025}.
However, it is suggestive that about half of the high-$z$ N-rich galaxies \tcrb{do} have AGN signatures.
In particular, GS\_3073, \tcra{the} galaxy with the highest N/O ratio reported so far at \tcra{$z\gtrsim5$} \citep{Ji2024}, is also a luminous Type-1 AGN \citep{Ubler2023}\tcra{, with clearly detected \tcrb{broad permitted line (hence BLR-originated) and} coronal lines that rule out photoionisation from stars alone \citep{Ji2024}}.
\tcrb{Conversely, the fact that we do not detect \niii] in the Non-AGN stack indicates that 
non-AGN galaxies as N-rich as those reported at $z\gtrsim5$ are a minority in the general galaxy population (while being a majority of the Type-1 AGN population).}

\section{Conclusions} \label{sec:con}
Using JWST/NIRSpec data obtained by the JADES survey, we find that the stacked spectrum of Type-1 AGNs at $z=4-7$ shows $\log(\mathrm{N/C})\simeq0.50$, $\log(\mathrm{N/O})>-0.58$, and $n_{\mathrm{e}}\sim10^{4}$\,cm$^{-3}$, which are higher than those of \tcrb{the Type-2 AGN and Non-AGN stacks} and comparable to those of high-$z$ N-rich galaxies.
Given that \tcrb{N enhancement is similarly} observed in \tcrb{GC}s, our findings suggest a close connection between dense proto-\tcrb{GC}s and black hole seeding in the early Universe. Specifically,
dense nuclear star formation may trap N-rich gas in proto-\tcrb{GC}s; associated runaway stellar collisions could produce intermediate-mass black hole seeds, whose accretion is visible through their broad line region.

\section*{Acknowledgements}

We thank S. Monty for having useful discussions of GCs.
YI is supported by JSPS KAKENHI Grant No. 24KJ0202.
RM, FDE, XJ, and JS acknowledge support by the Science and Technology Facilities Council (STFC), by the ERC through Advanced Grant 695671 ``QUENCH'', and by the UKRI Frontier Research grant RISEandFALL.
RM also acknowledges funding from a research professorship from the Royal Society.
%RM acknowledges support by the Science and Technology Facilities Council (STFC), by the ERC through Advanced Grant 695671 ``QUENCH'', and by the UKRI Frontier Research grant RISEandFALL. RM also acknowledges funding from a research professorship from the Royal Society.
%FDE acknowledges support by the Science and Technology Facilities Council (STFC), by the ERC through Advanced Grant 695671 ``QUENCH'', and by the UKRI Frontier Research grant RISEandFALL.
%XJ acknowledges support by the Science and Technology Facilities Council (STFC), by the ERC through Advanced Grant 695671 ``QUENCH'', and by the UKRI Frontier Research grant RISEandFALL.
IJ acknowledges support by the Huo Family Foundation through a P.C. Ho PhD Studentship.
%JS acknowledges support by the Science and Technology Facilities Council (STFC), ERC Advanced Grant 695671 ``QUENCH''.
AF acknowledges the support from project ``VLT- MOONS'' CRAM 1.05.03.07, INAF Large Grant 2022 ``The metal circle: a new sharp view of the baryon cycle up to Cosmic Dawn with the latest generation IFU facilities'' and INAF Large Grant 2022 ``Dual and binary SMBH in the multi-messenger era''.
%H\"U acknowledges support through the ERC Starting Grant 101164796 ``APEX''.
\tcre{H\"U acknowledges funding by the European Union (ERC APEX, 101164796). Views and opinions expressed are however those of the authors only and do not necessarily reflect those of the European Union or the European Research Council Executive Agency. Neither the European Union nor the granting authority can be held responsible for them.}
AJB acknowledges funding from the "FirstGalaxies" Advanced Grant from the European Research Council (ERC) under the European Union’s Horizon 2020 research and innovation programme (Grant agreement No. 789056).
SC acknowledges support by European Union's HE ERC Starting Grant No. 101040227 - WINGS.
ECL acknowledges support of an STFC Webb Fellowship (ST/W001438/1).
BER acknowledges support from the NIRCam Science Team contract to the University of Arizona, NAS5-02015, and JWST Program 3215.
JW gratefully acknowledges support from the Cosmic Dawn Center through the DAWN Fellowship. The Cosmic Dawn Center (DAWN) is funded by the Danish National Research Foundation under grant No. 140.
%This work is based on observations made with the NASA/ESA/CSA James Webb Space Telescope.
\tcrc{This work is based on observations made with the NASA/ESA/CSA James Webb Space Telescope. The data were obtained from the Mikulski Archive for Space Telescopes at the Space Telescope Science Institute, which is operated by the Association of Universities for Research in Astronomy, Inc., under NASA contract NAS 5-03127 for JWST. These observations are associated with programmes \#1180, 1181, 1210, 1286, 1287, and 3215.}
The authors acknowledge use of the lux supercomputer at UC Santa Cruz, funded by NSF MRI grant AST 1828315.
%The Acknowledgements section is not numbered. Here you can thank helpful colleagues, acknowledge funding agencies, telescopes and facilities used etc. Try to keep it short.

%%%%%%%%%%%%%%%%%%%%%%%%%%%%%%%%%%%%%%%%%%%%%%%%%%
\section*{Data Availability}

%The inclusion of a Data Availability Statement is a requirement for articles published in MNRAS. Data Availability Statements provide a standardised format for readers to understand the availability of data underlying the research results described in the article. The statement may refer to original data generated in the course of the study or to third-party data analysed in the article. The statement should describe and provide means of access, where possible, by linking to the data or providing the required accession numbers for the relevant databases or DOIs.
%The JWST/NIRSpec data used in this Letter are (or will be) available in the MAST archive.
\tcrb{The bulk of the JWST/NIRSpec data used in this Letter are released by the JADES NIRSpec DR1 \citep{Bunker2024} and DR3 \citep{DEugenio2025}, which are available on the JADES MAST website (\url{https://archive.stsci.edu/hlsp/jades}\tcre{; MAST DOI: \href{https://dx.doi.org/10.17909/8tdj-8n28}{10.17909/8tdj-8n28}}).
The rest of datasets will also be public in the MAST archive.}
Our analysed data will be made available upon reasonable request.



\bibliographystyle{mnras}
\bibliography{example} % if your bibtex file is called example.bib

\begin{thebibliography}{}
\makeatletter
\relax
\def\mn@urlcharsother{\let\do\@makeother \do\$\do\&\do\#\do\^\do\_\do\%\do\~}
\def\mn@doi{\begingroup\mn@urlcharsother \@ifnextchar [ {\mn@doi@} {\mn@doi@[]}}
\def\mn@doi@[#1]#2{\def\@tempa{#1}\ifx\@tempa\@empty \href {http://dx.doi.org/#2} {doi:#2}\else \href {http://dx.doi.org/#2} {#1}\fi \endgroup}
\def\mn@eprint#1#2{\mn@eprint@#1:#2::\@nil}
\def\mn@eprint@arXiv#1{\href {http://arxiv.org/abs/#1} {{\tt arXiv:#1}}}
\def\mn@eprint@dblp#1{\href {http://dblp.uni-trier.de/rec/bibtex/#1.xml} {dblp:#1}}
\def\mn@eprint@#1:#2:#3:#4\@nil{\def\@tempa {#1}\def\@tempb {#2}\def\@tempc {#3}\ifx \@tempc \@empty \let \@tempc \@tempb \let \@tempb \@tempa \fi \ifx \@tempb \@empty \def\@tempb {arXiv}\fi \@ifundefined {mn@eprint@\@tempb}{\@tempb:\@tempc}{\expandafter \expandafter \csname mn@eprint@\@tempb\endcsname \expandafter{\@tempc}}}

\bibitem[\protect\citeauthoryear{{{\'A}lvarez-M{\'a}rquez} et~al.,}{{{\'A}lvarez-M{\'a}rquez} et~al.}{2025}]{Alvarez2025}
{{\'A}lvarez-M{\'a}rquez} J.,  et~al., 2025, \mn@doi [\aap] {10.1051/0004-6361/202451731}, \href {https://ui.adsabs.harvard.edu/abs/2025A&A...695A.250A} {695, A250}

\bibitem[\protect\citeauthoryear{{Alves de Oliveira} et~al.,}{{Alves de Oliveira} et~al.}{2018}]{Oliveira2018}
{Alves de Oliveira} C.,  et~al., 2018, in Observatory Operations: Strategies, Processes, and Systems VII. p. 107040Q (\mn@eprint {arXiv} {1805.06922}), \mn@doi{10.1117/12.2313839}

\bibitem[\protect\citeauthoryear{{Amor{\'\i}n}, {P{\'e}rez-Montero}  \& {V{\'\i}lchez}}{{Amor{\'\i}n} et~al.}{2010}]{Amorin2010}
{Amor{\'\i}n} R.~O.,  {P{\'e}rez-Montero} E.,   {V{\'\i}lchez} J.~M.,  2010, \mn@doi [\apjl] {10.1088/2041-8205/715/2/L128}, \href {https://ui.adsabs.harvard.edu/abs/2010ApJ...715L.128A} {715, L128}

\bibitem[\protect\citeauthoryear{{Arellano-C{\'o}rdova} et~al.,}{{Arellano-C{\'o}rdova} et~al.}{2024}]{Arellano2024}
{Arellano-C{\'o}rdova} K.~Z.,  et~al., 2024, \mn@doi [arXiv e-prints] {10.48550/arXiv.2412.10557}, \href {https://ui.adsabs.harvard.edu/abs/2024arXiv241210557A} {p. arXiv:2412.10557}

\bibitem[\protect\citeauthoryear{{Asplund}, {Amarsi}  \& {Grevesse}}{{Asplund} et~al.}{2021}]{Asplund2021}
{Asplund} M.,  {Amarsi} A.~M.,   {Grevesse} N.,  2021, \mn@doi [\aap] {10.1051/0004-6361/202140445}, \href {https://ui.adsabs.harvard.edu/abs/2021A&A...653A.141A} {653, A141}

\bibitem[\protect\citeauthoryear{Baldwin, Phillips  \& Terlevich}{Baldwin et~al.}{1981}]{Baldwin1981}
Baldwin A.,  Phillips M.~M.,   Terlevich R.,  1981, \mn@doi [PASP] {10.1086/130930}, 93, 817

\bibitem[\protect\citeauthoryear{{Batra} \& {Baldwin}}{{Batra} \& {Baldwin}}{2014}]{Batra2014}
{Batra} N.~D.,  {Baldwin} J.~A.,  2014, \mn@doi [\mnras] {10.1093/mnras/stu007}, \href {https://ui.adsabs.harvard.edu/abs/2014MNRAS.439..771B} {439, 771}

\bibitem[\protect\citeauthoryear{{Berg}, {Skillman}, {Henry}, {Erb}  \& {Carigi}}{{Berg} et~al.}{2016}]{Berg2016}
{Berg} D.~A.,  {Skillman} E.~D.,  {Henry} R. B.~C.,  {Erb} D.~K.,   {Carigi} L.,  2016, \mn@doi [\apj] {10.3847/0004-637X/827/2/126}, \href {https://ui.adsabs.harvard.edu/abs/2016ApJ...827..126B} {827, 126}

\bibitem[\protect\citeauthoryear{Berg, Erb, Henry, Skillman  \& McQuinn}{Berg et~al.}{2019}]{Berg2019}
Berg D.~A.,  Erb D.~K.,  Henry R. B.~C.,  Skillman E.~D.,   McQuinn K. B.~W.,  2019, \mn@doi [ApJ] {10.3847/1538-4357/ab020a}, 874, 93

\bibitem[\protect\citeauthoryear{{B{\"o}ker} et~al.,}{{B{\"o}ker} et~al.}{2022}]{Boker2022}
{B{\"o}ker} T.,  et~al., 2022, \mn@doi [\aap] {10.1051/0004-6361/202142589}, \href {https://ui.adsabs.harvard.edu/abs/2022A&A...661A..82B} {661, A82}

\bibitem[\protect\citeauthoryear{{Bottorff}, {Lamothe}, {Momjian}, {Verner}, {Vinkovi{\'c}}  \& {Ferland}}{{Bottorff} et~al.}{1998}]{Bottorff1998}
{Bottorff} M.,  {Lamothe} J.,  {Momjian} E.,  {Verner} E.,  {Vinkovi{\'c}} D.,   {Ferland} G.,  1998, \mn@doi [\pasp] {10.1086/316222}, \href {https://ui.adsabs.harvard.edu/abs/1998PASP..110.1040B} {110, 1040}

\bibitem[\protect\citeauthoryear{{Bunker}, {NIRSPEC Instrument Science Team}  \& {JADESs Collaboration}}{{Bunker} et~al.}{2020}]{Bunker2020}
{Bunker} A.~J.,  {NIRSPEC Instrument Science Team}  {JADESs Collaboration} 2020, in {da Cunha} E.,  {Hodge} J.,  {Afonso} J.,  {Pentericci} L.,   {Sobral} D.,  eds,  IAU Symposium Vol. 352, Uncovering Early Galaxy Evolution in the ALMA and JWST Era. pp 342--346 (\mn@eprint {arXiv} {2112.15207}), \mn@doi{10.1017/S1743921319009463}

\bibitem[\protect\citeauthoryear{{Bunker} et~al.,}{{Bunker} et~al.}{2023}]{Bunker2023}
{Bunker} A.~J.,  et~al., 2023, \mn@doi [\aap] {10.1051/0004-6361/202346159}, \href {https://ui.adsabs.harvard.edu/abs/2023A&A...677A..88B} {677, A88}

\bibitem[\protect\citeauthoryear{{Bunker} et~al.,}{{Bunker} et~al.}{2024}]{Bunker2024}
{Bunker} A.~J.,  et~al., 2024, \mn@doi [\aap] {10.1051/0004-6361/202347094}, \href {https://ui.adsabs.harvard.edu/abs/2024A&A...690A.288B} {690, A288}

\bibitem[\protect\citeauthoryear{Calzetti, Armus, Bohlin, Kinney, Koornneef  \& Storchi‐Bergmann}{Calzetti et~al.}{2000}]{Calzetti2000}
Calzetti D.,  Armus L.,  Bohlin R.~C.,  Kinney A.~L.,  Koornneef J.,   Storchi‐Bergmann T.,  2000, \mn@doi [ApJ] {10.1086/308692}, 533, 682

\bibitem[\protect\citeauthoryear{{Cameron}, {Katz}, {Rey}  \& {Saxena}}{{Cameron} et~al.}{2023a}]{Cameron2023}
{Cameron} A.~J.,  {Katz} H.,  {Rey} M.~P.,   {Saxena} A.,  2023a, \mn@doi [\mnras] {10.1093/mnras/stad1579}, \href {https://ui.adsabs.harvard.edu/abs/2023MNRAS.523.3516C} {523, 3516}

\bibitem[\protect\citeauthoryear{{Cameron} et~al.,}{{Cameron} et~al.}{2023b}]{Cameron2023b}
{Cameron} A.~J.,  et~al., 2023b, \mn@doi [\aap] {10.1051/0004-6361/202346107}, \href {https://ui.adsabs.harvard.edu/abs/2023A&A...677A.115C} {677, A115}

\bibitem[\protect\citeauthoryear{{Cantiello}, {Jermyn}  \& {Lin}}{{Cantiello} et~al.}{2021}]{Cantiello2021}
{Cantiello} M.,  {Jermyn} A.~S.,   {Lin} D. N.~C.,  2021, \mn@doi [\apj] {10.3847/1538-4357/abdf4f}, \href {https://ui.adsabs.harvard.edu/abs/2021ApJ...910...94C} {910, 94}

\bibitem[\protect\citeauthoryear{{Carnall}}{{Carnall}}{2017}]{Carnall2017}
{Carnall} A.~C.,  2017, \mn@doi [arXiv e-prints] {10.48550/arXiv.1705.05165}, \href {https://ui.adsabs.harvard.edu/abs/2017arXiv170505165C} {p. arXiv:1705.05165}

\bibitem[\protect\citeauthoryear{{Carretta}, {Gratton}, {Lucatello}, {Bragaglia}  \& {Bonifacio}}{{Carretta} et~al.}{2005}]{Carretta2005}
{Carretta} E.,  {Gratton} R.~G.,  {Lucatello} S.,  {Bragaglia} A.,   {Bonifacio} P.,  2005, \mn@doi [\aap] {10.1051/0004-6361:20041892}, \href {https://ui.adsabs.harvard.edu/abs/2005A&A...433..597C} {433, 597}

\bibitem[\protect\citeauthoryear{{Castellano} et~al.,}{{Castellano} et~al.}{2024}]{Castellano2024}
{Castellano} M.,  et~al., 2024, \mn@doi [\apj] {10.3847/1538-4357/ad5f88}, \href {https://ui.adsabs.harvard.edu/abs/2024ApJ...972..143C} {972, 143}

\bibitem[\protect\citeauthoryear{{Charbonnel}, {Schaerer}, {Prantzos}, {Ram{\'\i}rez-Galeano}, {Fragos}, {Kuruvanthodi}, {Marques-Chaves}  \& {Gieles}}{{Charbonnel} et~al.}{2023}]{Charbonnel2023}
{Charbonnel} C.,  {Schaerer} D.,  {Prantzos} N.,  {Ram{\'\i}rez-Galeano} L.,  {Fragos} T.,  {Kuruvanthodi} A.,  {Marques-Chaves} R.,   {Gieles} M.,  2023, \mn@doi [\aap] {10.1051/0004-6361/202346410}, \href {https://ui.adsabs.harvard.edu/abs/2023A&A...673L...7C} {673, L7}

\bibitem[\protect\citeauthoryear{{Curti} et~al.,}{{Curti} et~al.}{2024}]{Curti2024}
{Curti} M.,  et~al., 2024, \mn@doi [arXiv e-prints] {10.48550/arXiv.2407.02575}, \href {https://ui.adsabs.harvard.edu/abs/2024arXiv240702575C} {p. arXiv:2407.02575}

\bibitem[\protect\citeauthoryear{{D'Antona} et~al.,}{{D'Antona} et~al.}{2023}]{DAntona2023}
{D'Antona} F.,  et~al., 2023, \mn@doi [\aap] {10.1051/0004-6361/202348240}, \href {https://ui.adsabs.harvard.edu/abs/2023A&A...680L..19D} {680, L19}

\bibitem[\protect\citeauthoryear{{D'Eugenio} et~al.,}{{D'Eugenio} et~al.}{2025}]{DEugenio2025}
{D'Eugenio} F.,  et~al., 2025, \mn@doi [\apjs] {10.3847/1538-4365/ada148}, \href {https://ui.adsabs.harvard.edu/abs/2025ApJS..277....4D} {277, 4}

\bibitem[\protect\citeauthoryear{{Dall'Amico}, {Mapelli}, {Iorio}, {Costa}, {Charlot}, {Korb}, {Sgalletta}  \& {Lecroq}}{{Dall'Amico} et~al.}{2025}]{DallAmico2025}
{Dall'Amico} M.,  {Mapelli} M.,  {Iorio} G.,  {Costa} G.,  {Charlot} S.,  {Korb} E.,  {Sgalletta} C.,   {Lecroq} M.,  2025, \mn@doi [\aap] {10.1051/0004-6361/202453543}, \href {https://ui.adsabs.harvard.edu/abs/2025A&A...695A.221D} {695, A221}

\bibitem[\protect\citeauthoryear{{Dorner} et~al.,}{{Dorner} et~al.}{2016}]{Dorner2016}
{Dorner} B.,  et~al., 2016, \mn@doi [\aap] {10.1051/0004-6361/201628263}, \href {https://ui.adsabs.harvard.edu/abs/2016A&A...592A.113D} {592, A113}

\bibitem[\protect\citeauthoryear{{Eisenstein} et~al.,}{{Eisenstein} et~al.}{2023a}]{Eisenstein2023}
{Eisenstein} D.~J.,  et~al., 2023a, \mn@doi [arXiv e-prints] {10.48550/arXiv.2306.02465}, \href {https://ui.adsabs.harvard.edu/abs/2023arXiv230602465E} {p. arXiv:2306.02465}

\bibitem[\protect\citeauthoryear{{Eisenstein} et~al.,}{{Eisenstein} et~al.}{2023b}]{Eisenstein2023b}
{Eisenstein} D.~J.,  et~al., 2023b, \mn@doi [arXiv e-prints] {10.48550/arXiv.2310.12340}, \href {https://ui.adsabs.harvard.edu/abs/2023arXiv231012340E} {p. arXiv:2310.12340}

\bibitem[\protect\citeauthoryear{{Ferland} et~al.,}{{Ferland} et~al.}{2013}]{Ferland2013}
{Ferland} G.~J.,  et~al., 2013, \rmxaa, \href {https://ui.adsabs.harvard.edu/abs/2013RMxAA..49..137F} {49, 137}

\bibitem[\protect\citeauthoryear{{Ferruit} et~al.,}{{Ferruit} et~al.}{2022}]{Ferruit2022}
{Ferruit} P.,  et~al., 2022, \mn@doi [\aap] {10.1051/0004-6361/202142673}, \href {https://ui.adsabs.harvard.edu/abs/2022A&A...661A..81F} {661, A81}

\bibitem[\protect\citeauthoryear{{Fukushima} \& {Yajima}}{{Fukushima} \& {Yajima}}{2024}]{Fukushima2024}
{Fukushima} H.,  {Yajima} H.,  2024, \mn@doi [\pasj] {10.1093/pasj/psae074}, \href {https://ui.adsabs.harvard.edu/abs/2024PASJ...76.1122F} {76, 1122}

\bibitem[\protect\citeauthoryear{{Gardner} et~al.,}{{Gardner} et~al.}{2023}]{Gardner2023}
{Gardner} J.~P.,  et~al., 2023, \mn@doi [\pasp] {10.1088/1538-3873/acd1b5}, \href {https://ui.adsabs.harvard.edu/abs/2023PASP..135f8001G} {135, 068001}

\bibitem[\protect\citeauthoryear{{Garnett}}{{Garnett}}{1992}]{Garnett1992}
{Garnett} D.~R.,  1992, \mn@doi [\aj] {10.1086/116146}, \href {https://ui.adsabs.harvard.edu/abs/1992AJ....103.1330G} {103, 1330}

\bibitem[\protect\citeauthoryear{{Gieles} et~al.,}{{Gieles} et~al.}{2018}]{Gieles2018}
{Gieles} M.,  et~al., 2018, \mn@doi [\mnras] {10.1093/mnras/sty1059}, \href {https://ui.adsabs.harvard.edu/abs/2018MNRAS.478.2461G} {478, 2461}

\bibitem[\protect\citeauthoryear{{Gordon}, {Clayton}, {Misselt}, {Landolt}  \& {Wolff}}{{Gordon} et~al.}{2003}]{Gordon2003}
{Gordon} K.~D.,  {Clayton} G.~C.,  {Misselt} K.~A.,  {Landolt} A.~U.,   {Wolff} M.~J.,  2003, \mn@doi [\apj] {10.1086/376774}, \href {https://ui.adsabs.harvard.edu/abs/2003ApJ...594..279G} {594, 279}

\bibitem[\protect\citeauthoryear{{H{\"a}berle} et~al.,}{{H{\"a}berle} et~al.}{2024}]{Haberle2024}
{H{\"a}berle} M.,  et~al., 2024, \mn@doi [\nat] {10.1038/s41586-024-07511-z}, \href {https://ui.adsabs.harvard.edu/abs/2024Natur.631..285H} {631, 285}

\bibitem[\protect\citeauthoryear{{Harikane} et~al.,}{{Harikane} et~al.}{2023}]{Harikane2023}
{Harikane} Y.,  et~al., 2023, \mn@doi [\apj] {10.3847/1538-4357/ad029e}, \href {https://ui.adsabs.harvard.edu/abs/2023ApJ...959...39H} {959, 39}

\bibitem[\protect\citeauthoryear{{Harikane} et~al.,}{{Harikane} et~al.}{2025}]{Harikane2025}
{Harikane} Y.,  et~al., 2025, \mn@doi [\apj] {10.3847/1538-4357/ad9b2c}, \href {https://ui.adsabs.harvard.edu/abs/2025ApJ...980..138H} {980, 138}

\bibitem[\protect\citeauthoryear{{Hayes}, {Saldana-Lopez}, {Citro}, {James}, {Mingozzi}, {Scarlata}, {Martinez}  \& {Berg}}{{Hayes} et~al.}{2025}]{Hayes2025}
{Hayes} M.~J.,  {Saldana-Lopez} A.,  {Citro} A.,  {James} B.~L.,  {Mingozzi} M.,  {Scarlata} C.,  {Martinez} Z.,   {Berg} D.~A.,  2025, \mn@doi [\apj] {10.3847/1538-4357/adaea1}, \href {https://ui.adsabs.harvard.edu/abs/2025ApJ...982...14H} {982, 14}

\bibitem[\protect\citeauthoryear{{Heard} \& {Gaskell}}{{Heard} \& {Gaskell}}{2016}]{Heard2016}
{Heard} C. Z.~P.,  {Gaskell} C.~M.,  2016, \mn@doi [\mnras] {10.1093/mnras/stw1616}, \href {https://ui.adsabs.harvard.edu/abs/2016MNRAS.461.4227H} {461, 4227}

\bibitem[\protect\citeauthoryear{{Hirschmann} et~al.,}{{Hirschmann} et~al.}{2023}]{Hirschmann2023}
{Hirschmann} M.,  et~al., 2023, \mn@doi [\mnras] {10.1093/mnras/stad2955}, \href {https://ui.adsabs.harvard.edu/abs/2023MNRAS.526.3610H} {526, 3610}

\bibitem[\protect\citeauthoryear{{Inayoshi}, {Visbal}  \& {Haiman}}{{Inayoshi} et~al.}{2020}]{Inayoshi2020}
{Inayoshi} K.,  {Visbal} E.,   {Haiman} Z.,  2020, \mn@doi [\araa] {10.1146/annurev-astro-120419-014455}, \href {https://ui.adsabs.harvard.edu/abs/2020ARA&A..58...27I} {58, 27}

\bibitem[\protect\citeauthoryear{{Inoguchi}, {Hosokawa}, {Fukushima}, {Tanaka}, {Yajima}  \& {Mineshige}}{{Inoguchi} et~al.}{2024}]{Inoguchi2024}
{Inoguchi} M.,  {Hosokawa} T.,  {Fukushima} H.,  {Tanaka} K. E.~I.,  {Yajima} H.,   {Mineshige} S.,  2024, \mn@doi [\mnras] {10.1093/mnras/stad3297}, \href {https://ui.adsabs.harvard.edu/abs/2024MNRAS.527.3612I} {527, 3612}

\bibitem[\protect\citeauthoryear{{Isobe} et~al.,}{{Isobe} et~al.}{2022}]{Isobe2022}
{Isobe} Y.,  et~al., 2022, \mn@doi [\apj] {10.3847/1538-4357/ac3509}, \href {https://ui.adsabs.harvard.edu/abs/2022ApJ...925..111I} {925, 111}

\bibitem[\protect\citeauthoryear{{Isobe}, {Ouchi}, {Nakajima}, {Harikane}, {Ono}, {Xu}, {Zhang}  \& {Umeda}}{{Isobe} et~al.}{2023a}]{Isobe2023b}
{Isobe} Y.,  {Ouchi} M.,  {Nakajima} K.,  {Harikane} Y.,  {Ono} Y.,  {Xu} Y.,  {Zhang} Y.,   {Umeda} H.,  2023a, \mn@doi [\apj] {10.3847/1538-4357/acf376}, \href {https://ui.adsabs.harvard.edu/abs/2023ApJ...956..139I} {956, 139}

\bibitem[\protect\citeauthoryear{{Isobe} et~al.,}{{Isobe} et~al.}{2023b}]{Isobe2023c}
{Isobe} Y.,  et~al., 2023b, \mn@doi [\apj] {10.3847/1538-4357/ad09be}, \href {https://ui.adsabs.harvard.edu/abs/2023ApJ...959..100I} {959, 100}

\bibitem[\protect\citeauthoryear{Izotov, Stasi{\'{n}}ska, Meynet, Guseva  \& Thuan}{Izotov et~al.}{2006}]{Izotov2006}
Izotov Y.~I.,  Stasi{\'{n}}ska G.,  Meynet G.,  Guseva N.~G.,   Thuan T.~X.,  2006, \mn@doi [A{\&}A] {10.1051/0004-6361:20053763}, 448, 955

\bibitem[\protect\citeauthoryear{{Jakobsen} et~al.,}{{Jakobsen} et~al.}{2022}]{Jakobsen2022}
{Jakobsen} P.,  et~al., 2022, \mn@doi [\aap] {10.1051/0004-6361/202142663}, \href {https://ui.adsabs.harvard.edu/abs/2022A&A...661A..80J} {661, A80}

\bibitem[\protect\citeauthoryear{{Ji} et~al.,}{{Ji} et~al.}{2024}]{Ji2024}
{Ji} X.,  et~al., 2024, \mn@doi [\mnras] {10.1093/mnras/stae2375}, \href {https://ui.adsabs.harvard.edu/abs/2024MNRAS.535..881J} {535, 881}

\bibitem[\protect\citeauthoryear{{Juod{\v{z}}balis} et~al.,}{{Juod{\v{z}}balis} et~al.}{2024}]{Juodzbalis2024}
{Juod{\v{z}}balis} I.,  et~al., 2024, \mn@doi [\nat] {10.1038/s41586-024-08210-5}, \href {https://ui.adsabs.harvard.edu/abs/2024Natur.636..594J} {636, 594}

\bibitem[\protect\citeauthoryear{{Juod{\v{z}}balis} et~al.,}{{Juod{\v{z}}balis} et~al.}{2025}]{Juodzbalis2025}
{Juod{\v{z}}balis} I.,  et~al., 2025, \mn@doi [arXiv e-prints] {10.48550/arXiv.2504.03551}, \href {https://ui.adsabs.harvard.edu/abs/2025arXiv250403551J} {p. arXiv:2504.03551}

\bibitem[\protect\citeauthoryear{{Kobayashi} \& {Ferrara}}{{Kobayashi} \& {Ferrara}}{2024}]{Kobayashi2024}
{Kobayashi} C.,  {Ferrara} A.,  2024, \mn@doi [\apjl] {10.3847/2041-8213/ad1de1}, \href {https://ui.adsabs.harvard.edu/abs/2024ApJ...962L...6K} {962, L6}

\bibitem[\protect\citeauthoryear{{Kocevski} et~al.,}{{Kocevski} et~al.}{2023}]{Kocevski2023}
{Kocevski} D.~D.,  et~al., 2023, \mn@doi [\apjl] {10.3847/2041-8213/ace5a0}, \href {https://ui.adsabs.harvard.edu/abs/2023ApJ...954L...4K} {954, L4}

\bibitem[\protect\citeauthoryear{{Kojima} et~al.,}{{Kojima} et~al.}{2020}]{Kojima2020}
{Kojima} T.,  et~al., 2020, \mn@doi [\apj] {10.3847/1538-4357/aba047}, \href {https://ui.adsabs.harvard.edu/abs/2020ApJ...898..142K} {898, 142}

\bibitem[\protect\citeauthoryear{{Kumari}, {James}, {Irwin}, {Amor{\'\i}n}  \& {P{\'e}rez-Montero}}{{Kumari} et~al.}{2018}]{Kumari2018}
{Kumari} N.,  {James} B.~L.,  {Irwin} M.~J.,  {Amor{\'\i}n} R.,   {P{\'e}rez-Montero} E.,  2018, \mn@doi [\mnras] {10.1093/mnras/sty402}, \href {https://ui.adsabs.harvard.edu/abs/2018MNRAS.476.3793K} {476, 3793}

\bibitem[\protect\citeauthoryear{{Lambrides} et~al.,}{{Lambrides} et~al.}{2024}]{Lambrides2024}
{Lambrides} E.,  et~al., 2024, \mn@doi [arXiv e-prints] {10.48550/arXiv.2409.13047}, \href {https://ui.adsabs.harvard.edu/abs/2024arXiv240913047L} {p. arXiv:2409.13047}

\bibitem[\protect\citeauthoryear{{Larson} et~al.,}{{Larson} et~al.}{2023}]{Larson2023}
{Larson} R.~L.,  et~al., 2023, \mn@doi [\apjl] {10.3847/2041-8213/ace619}, \href {https://ui.adsabs.harvard.edu/abs/2023ApJ...953L..29L} {953, L29}

\bibitem[\protect\citeauthoryear{{Lennon} \& {Burke}}{{Lennon} \& {Burke}}{1994}]{Lennon1994}
{Lennon} D.~J.,  {Burke} V.~M.,  1994, \aaps, \href {https://ui.adsabs.harvard.edu/abs/1994A&AS..103..273L} {103, 273}

\bibitem[\protect\citeauthoryear{{Li} et~al.,}{{Li} et~al.}{2025}]{Li2025}
{Li} S.,  et~al., 2025, \mn@doi [\apjl] {10.3847/2041-8213/ad9eac}, \href {https://ui.adsabs.harvard.edu/abs/2025ApJ...979L..13L} {979, L13}

\bibitem[\protect\citeauthoryear{{Luridiana}, {Morisset}  \& {Shaw}}{{Luridiana} et~al.}{2015}]{Luridiana2015}
{Luridiana} V.,  {Morisset} C.,   {Shaw} R.~A.,  2015, \mn@doi [\aap] {10.1051/0004-6361/201323152}, \href {https://ui.adsabs.harvard.edu/abs/2015A&A...573A..42L} {573, A42}

\bibitem[\protect\citeauthoryear{{Maiolino} et~al.,}{{Maiolino} et~al.}{2024a}]{Maiolino2024a}
{Maiolino} R.,  et~al., 2024a, \mn@doi [\nat] {10.1038/s41586-024-07052-5}, \href {https://ui.adsabs.harvard.edu/abs/2024Natur.627...59M} {627, 59}

\bibitem[\protect\citeauthoryear{{Maiolino} et~al.,}{{Maiolino} et~al.}{2024b}]{Maiolino2024c}
{Maiolino} R.,  et~al., 2024b, \mn@doi [\aap] {10.1051/0004-6361/202347640}, \href {https://ui.adsabs.harvard.edu/abs/2024A&A...691A.145M} {691, A145}

\bibitem[\protect\citeauthoryear{{Marques-Chaves} et~al.,}{{Marques-Chaves} et~al.}{2024}]{Marques-Chaves2024}
{Marques-Chaves} R.,  et~al., 2024, \mn@doi [\aap] {10.1051/0004-6361/202347411}, \href {https://ui.adsabs.harvard.edu/abs/2024A&A...681A..30M} {681, A30}

\bibitem[\protect\citeauthoryear{{Mazzolari} et~al.,}{{Mazzolari} et~al.}{2024}]{Mazzolari2024}
{Mazzolari} G.,  et~al., 2024, \mn@doi [\aap] {10.1051/0004-6361/202450407}, \href {https://ui.adsabs.harvard.edu/abs/2024A&A...691A.345M} {691, A345}

\bibitem[\protect\citeauthoryear{{Nagele} \& {Umeda}}{{Nagele} \& {Umeda}}{2023}]{Nagele2023}
{Nagele} C.,  {Umeda} H.,  2023, \mn@doi [\apjl] {10.3847/2041-8213/acd550}, \href {https://ui.adsabs.harvard.edu/abs/2023ApJ...949L..16N} {949, L16}

\bibitem[\protect\citeauthoryear{{Nandal}, {Regan}, {Woods}, {Farrell}, {Ekstr{\"o}m}  \& {Meynet}}{{Nandal} et~al.}{2024}]{Nandal2024a}
{Nandal} D.,  {Regan} J.~A.,  {Woods} T.~E.,  {Farrell} E.,  {Ekstr{\"o}m} S.,   {Meynet} G.,  2024, \mn@doi [\aap] {10.1051/0004-6361/202348035}, \href {https://ui.adsabs.harvard.edu/abs/2024A&A...683A.156N} {683, A156}

\bibitem[\protect\citeauthoryear{{Napolitano} et~al.,}{{Napolitano} et~al.}{2024}]{Napolitano2024b}
{Napolitano} L.,  et~al., 2024, \mn@doi [arXiv e-prints] {10.48550/arXiv.2410.18763}, \href {https://ui.adsabs.harvard.edu/abs/2024arXiv241018763N} {p. arXiv:2410.18763}

\bibitem[\protect\citeauthoryear{{Navarro-Carrera}, {Caputi}, {Iani}, {Rinaldi}, {Kokorev}  \& {Kerutt}}{{Navarro-Carrera} et~al.}{2024}]{Navarro-Carrera2024}
{Navarro-Carrera} R.,  {Caputi} K.~I.,  {Iani} E.,  {Rinaldi} P.,  {Kokorev} V.,   {Kerutt} J.,  2024, \mn@doi [arXiv e-prints] {10.48550/arXiv.2407.14201}, \href {https://ui.adsabs.harvard.edu/abs/2024arXiv240714201N} {p. arXiv:2407.14201}

\bibitem[\protect\citeauthoryear{{Nicholls}, {Sutherland}, {Dopita}, {Kewley}  \& {Groves}}{{Nicholls} et~al.}{2017}]{Nicholls2017}
{Nicholls} D.~C.,  {Sutherland} R.~S.,  {Dopita} M.~A.,  {Kewley} L.~J.,   {Groves} B.~A.,  2017, \mn@doi [\mnras] {10.1093/mnras/stw3235}, \href {https://ui.adsabs.harvard.edu/abs/2017MNRAS.466.4403N} {466, 4403}

\bibitem[\protect\citeauthoryear{{Partmann}, {Naab}, {Lah{\'e}n}, {Rantala}, {Hirschmann}, {Hislop}, {Petersson}  \& {Johansson}}{{Partmann} et~al.}{2025}]{Partmann2025}
{Partmann} C.,  {Naab} T.,  {Lah{\'e}n} N.,  {Rantala} A.,  {Hirschmann} M.,  {Hislop} J.~M.,  {Petersson} J.,   {Johansson} P.~H.,  2025, \mn@doi [\mnras] {10.1093/mnras/staf002}, \href {https://ui.adsabs.harvard.edu/abs/2025MNRAS.537..956P} {537, 956}

\bibitem[\protect\citeauthoryear{{Pascale}, {Dai}, {McKee}  \& {Tsang}}{{Pascale} et~al.}{2023}]{Pascale2023}
{Pascale} M.,  {Dai} L.,  {McKee} C.~F.,   {Tsang} B. T.~H.,  2023, \mn@doi [\apj] {10.3847/1538-4357/acf75c}, \href {https://ui.adsabs.harvard.edu/abs/2023ApJ...957...77P} {957, 77}

\bibitem[\protect\citeauthoryear{{Rantala}, {Naab}  \& {Lah{\'e}n}}{{Rantala} et~al.}{2024a}]{Rantala2024}
{Rantala} A.,  {Naab} T.,   {Lah{\'e}n} N.,  2024a, \mn@doi [\mnras] {10.1093/mnras/stae1413}, \href {https://ui.adsabs.harvard.edu/abs/2024MNRAS.531.3770R} {531, 3770}

\bibitem[\protect\citeauthoryear{{Rantala}, {Rawlings}, {Naab}, {Thomas}  \& {Johansson}}{{Rantala} et~al.}{2024b}]{Antti2024}
{Rantala} A.,  {Rawlings} A.,  {Naab} T.,  {Thomas} J.,   {Johansson} P.~H.,  2024b, \mn@doi [\mnras] {10.1093/mnras/stae2424}, \href {https://ui.adsabs.harvard.edu/abs/2024MNRAS.535.1202R} {535, 1202}

\bibitem[\protect\citeauthoryear{{Reddy}, {Topping}, {Sanders}, {Shapley}  \& {Brammer}}{{Reddy} et~al.}{2023a}]{Reddy2023a}
{Reddy} N.~A.,  {Topping} M.~W.,  {Sanders} R.~L.,  {Shapley} A.~E.,   {Brammer} G.,  2023a, \mn@doi [\apj] {10.3847/1538-4357/acc869}, \href {https://ui.adsabs.harvard.edu/abs/2023ApJ...948...83R} {948, 83}

\bibitem[\protect\citeauthoryear{{Reddy}, {Topping}, {Sanders}, {Shapley}  \& {Brammer}}{{Reddy} et~al.}{2023b}]{Reddy2023b}
{Reddy} N.~A.,  {Topping} M.~W.,  {Sanders} R.~L.,  {Shapley} A.~E.,   {Brammer} G.,  2023b, \mn@doi [\apj] {10.3847/1538-4357/acd754}, \href {https://ui.adsabs.harvard.edu/abs/2023ApJ...952..167R} {952, 167}

\bibitem[\protect\citeauthoryear{{Renzini}}{{Renzini}}{2023}]{Renzini2023}
{Renzini} A.,  2023, \mn@doi [\mnras] {10.1093/mnrasl/slad091}, \href {https://ui.adsabs.harvard.edu/abs/2023MNRAS.525L.117R} {525, L117}

\bibitem[\protect\citeauthoryear{{Rieke}}{{Rieke}}{2020}]{Rieke2020}
{Rieke} M.,  2020, in {da Cunha} E.,  {Hodge} J.,  {Afonso} J.,  {Pentericci} L.,   {Sobral} D.,  eds,  IAU Symposium Vol. 352, Uncovering Early Galaxy Evolution in the ALMA and JWST Era. pp 337--341, \mn@doi{10.1017/S1743921319008950}

\bibitem[\protect\citeauthoryear{{Rigby} et~al.,}{{Rigby} et~al.}{2023}]{Rigby2023}
{Rigby} J.,  et~al., 2023, \mn@doi [\pasp] {10.1088/1538-3873/acb293}, \href {https://ui.adsabs.harvard.edu/abs/2023PASP..135d8001R} {135, 048001}

\bibitem[\protect\citeauthoryear{{Rizzuti}, {Matteucci}, {Molaro}, {Cescutti}  \& {Maiolino}}{{Rizzuti} et~al.}{2024}]{Rizzuti2024}
{Rizzuti} F.,  {Matteucci} F.,  {Molaro} P.,  {Cescutti} G.,   {Maiolino} R.,  2024, \mn@doi [arXiv e-prints] {10.48550/arXiv.2412.05363}, \href {https://ui.adsabs.harvard.edu/abs/2024arXiv241205363R} {p. arXiv:2412.05363}

\bibitem[\protect\citeauthoryear{{Sakurai}, {Yoshida}  \& {Fujii}}{{Sakurai} et~al.}{2019}]{Sakurai2019}
{Sakurai} Y.,  {Yoshida} N.,   {Fujii} M.~S.,  2019, \mn@doi [\mnras] {10.1093/mnras/stz315}, \href {https://ui.adsabs.harvard.edu/abs/2019MNRAS.484.4665S} {484, 4665}

\bibitem[\protect\citeauthoryear{{Salpeter}}{{Salpeter}}{1955}]{Salpeter1955}
{Salpeter} E.~E.,  1955, \mn@doi [\apj] {10.1086/145971}, \href {https://ui.adsabs.harvard.edu/abs/1955ApJ...121..161S} {121, 161}

\bibitem[\protect\citeauthoryear{{Sandles} et~al.,}{{Sandles} et~al.}{2024}]{Sandles2024}
{Sandles} L.,  et~al., 2024, \mn@doi [\aap] {10.1051/0004-6361/202347119}, \href {https://ui.adsabs.harvard.edu/abs/2024A&A...691A.305S} {691, A305}

\bibitem[\protect\citeauthoryear{{Schaerer}, {Marques-Chaves}, {Xiao}  \& {Korber}}{{Schaerer} et~al.}{2024}]{Schaerer2024}
{Schaerer} D.,  {Marques-Chaves} R.,  {Xiao} M.,   {Korber} D.,  2024, \mn@doi [\aap] {10.1051/0004-6361/202450721}, \href {https://ui.adsabs.harvard.edu/abs/2024A&A...687L..11S} {687, L11}

\bibitem[\protect\citeauthoryear{{Scholtz} et~al.,}{{Scholtz} et~al.}{2023}]{Scholtz2023}
{Scholtz} J.,  et~al., 2023, \mn@doi [arXiv e-prints] {10.48550/arXiv.2311.18731}, \href {https://ui.adsabs.harvard.edu/abs/2023arXiv231118731S} {p. arXiv:2311.18731}

\bibitem[\protect\citeauthoryear{{Senchyna}, {Plat}, {Stark}, {Rudie}, {Berg}, {Charlot}, {James}  \& {Mingozzi}}{{Senchyna} et~al.}{2024}]{Senchyna2024}
{Senchyna} P.,  {Plat} A.,  {Stark} D.~P.,  {Rudie} G.~C.,  {Berg} D.,  {Charlot} S.,  {James} B.~L.,   {Mingozzi} M.,  2024, \mn@doi [\apj] {10.3847/1538-4357/ad235e}, \href {https://ui.adsabs.harvard.edu/abs/2024ApJ...966...92S} {966, 92}

\bibitem[\protect\citeauthoryear{{Stanway} \& {Eldridge}}{{Stanway} \& {Eldridge}}{2018}]{Stanway2018}
{Stanway} E.~R.,  {Eldridge} J.~J.,  2018, \mn@doi [\mnras] {10.1093/mnras/sty1353}, \href {https://ui.adsabs.harvard.edu/abs/2018MNRAS.479...75S} {479, 75}

\bibitem[\protect\citeauthoryear{{Steidel}, {Strom}, {Pettini}, {Rudie}, {Reddy}  \& {Trainor}}{{Steidel} et~al.}{2016}]{Steidel2016}
{Steidel} C.~C.,  {Strom} A.~L.,  {Pettini} M.,  {Rudie} G.~C.,  {Reddy} N.~A.,   {Trainor} R.~F.,  2016, \mn@doi [\apj] {10.3847/0004-637X/826/2/159}, \href {https://ui.adsabs.harvard.edu/abs/2016ApJ...826..159S} {826, 159}

\bibitem[\protect\citeauthoryear{{Stiavelli} et~al.,}{{Stiavelli} et~al.}{2025}]{Stiavelli2025}
{Stiavelli} M.,  et~al., 2025, \mn@doi [\apj] {10.3847/1538-4357/adb5f3}, \href {https://ui.adsabs.harvard.edu/abs/2025ApJ...981..136S} {981, 136}

\bibitem[\protect\citeauthoryear{{Taylor} et~al.,}{{Taylor} et~al.}{2024}]{Taylor2024}
{Taylor} A.~J.,  et~al., 2024, \mn@doi [arXiv e-prints] {10.48550/arXiv.2409.06772}, \href {https://ui.adsabs.harvard.edu/abs/2024arXiv240906772T} {p. arXiv:2409.06772}

\bibitem[\protect\citeauthoryear{{Telles}, {Thuan}, {Izotov}  \& {Carrasco}}{{Telles} et~al.}{2014}]{Telles2014}
{Telles} E.,  {Thuan} T.~X.,  {Izotov} Y.~I.,   {Carrasco} E.~R.,  2014, \mn@doi [\aap] {10.1051/0004-6361/201219270}, \href {https://ui.adsabs.harvard.edu/abs/2014A&A...561A..64T} {561, A64}

\bibitem[\protect\citeauthoryear{{Temple}, {Ferland}, {Rankine}, {Chatzikos}  \& {Hewett}}{{Temple} et~al.}{2021}]{Temple2021}
{Temple} M.~J.,  {Ferland} G.~J.,  {Rankine} A.~L.,  {Chatzikos} M.,   {Hewett} P.~C.,  2021, \mn@doi [\mnras] {10.1093/mnras/stab1610}, \href {https://ui.adsabs.harvard.edu/abs/2021MNRAS.505.3247T} {505, 3247}

\bibitem[\protect\citeauthoryear{{Topping} et~al.,}{{Topping} et~al.}{2024}]{Topping2024a}
{Topping} M.~W.,  et~al., 2024, \mn@doi [\mnras] {10.1093/mnras/stae682}, \href {https://ui.adsabs.harvard.edu/abs/2024MNRAS.529.3301T} {529, 3301}

\bibitem[\protect\citeauthoryear{{Topping} et~al.,}{{Topping} et~al.}{2025}]{Topping2025}
{Topping} M.~W.,  et~al., 2025, \mn@doi [\apj] {10.3847/1538-4357/ada95c}, \href {https://ui.adsabs.harvard.edu/abs/2025ApJ...980..225T} {980, 225}

\bibitem[\protect\citeauthoryear{{Trinca} et~al.,}{{Trinca} et~al.}{2024}]{Trinca2024}
{Trinca} A.,  et~al., 2024, \mn@doi [arXiv e-prints] {10.48550/arXiv.2412.14248}, \href {https://ui.adsabs.harvard.edu/abs/2024arXiv241214248T} {p. arXiv:2412.14248}

\bibitem[\protect\citeauthoryear{{Tripodi} et~al.,}{{Tripodi} et~al.}{2024}]{Tripodi2024}
{Tripodi} R.,  et~al., 2024, \mn@doi [arXiv e-prints] {10.48550/arXiv.2412.04983}, \href {https://ui.adsabs.harvard.edu/abs/2024arXiv241204983T} {p. arXiv:2412.04983}

\bibitem[\protect\citeauthoryear{{{\"U}bler} et~al.,}{{{\"U}bler} et~al.}{2023}]{Ubler2023}
{{\"U}bler} H.,  et~al., 2023, \mn@doi [\aap] {10.1051/0004-6361/202346137}, \href {https://ui.adsabs.harvard.edu/abs/2023A&A...677A.145U} {677, A145}

\bibitem[\protect\citeauthoryear{{Veilleux} \& {Osterbrock}}{{Veilleux} \& {Osterbrock}}{1987}]{VO87}
{Veilleux} S.,  {Osterbrock} D.~E.,  1987, \mn@doi [\apjs] {10.1086/191166}, \href {https://ui.adsabs.harvard.edu/abs/1987ApJS...63..295V} {63, 295}

\bibitem[\protect\citeauthoryear{Vincenzo, Belfiore, Maiolino, Matteucci  \& Ventura}{Vincenzo et~al.}{2016}]{Vincenzo2016}
Vincenzo F.,  Belfiore F.,  Maiolino R.,  Matteucci F.,   Ventura P.,  2016, \mn@doi [MNRAS] {10.1093/mnras/stw532}, 458, 3466

\bibitem[\protect\citeauthoryear{{Vink}}{{Vink}}{2023}]{Vink2023}
{Vink} J.~S.,  2023, \mn@doi [\aap] {10.1051/0004-6361/202347827}, \href {https://ui.adsabs.harvard.edu/abs/2023A&A...679L...9V} {679, L9}

\bibitem[\protect\citeauthoryear{{Watanabe} et~al.,}{{Watanabe} et~al.}{2024}]{Watanabe2024}
{Watanabe} K.,  et~al., 2024, \mn@doi [\apj] {10.3847/1538-4357/ad13ff}, \href {https://ui.adsabs.harvard.edu/abs/2024ApJ...962...50W} {962, 50}

\bibitem[\protect\citeauthoryear{{Yanagisawa} et~al.,}{{Yanagisawa} et~al.}{2024}]{Yanagisawa2024}
{Yanagisawa} H.,  et~al., 2024, \mn@doi [\apj] {10.3847/1538-4357/ad72ec}, \href {https://ui.adsabs.harvard.edu/abs/2024ApJ...974..266Y} {974, 266}

\bibitem[\protect\citeauthoryear{{Zhang}, {Morishita}  \& {Stiavelli}}{{Zhang} et~al.}{2025}]{Zhang2025}
{Zhang} Y.,  {Morishita} T.,   {Stiavelli} M.,  2025, arXiv e-prints, \href {https://ui.adsabs.harvard.edu/abs/2025arXiv250204817Z} {p. arXiv:2502.04817}

\bibitem[\protect\citeauthoryear{{de Graaff} et~al.,}{{de Graaff} et~al.}{2024}]{deGraaff2024}
{de Graaff} A.,  et~al., 2024, \mn@doi [\aap] {10.1051/0004-6361/202347755}, \href {https://ui.adsabs.harvard.edu/abs/2024A&A...684A..87D} {684, A87}

\makeatother
\end{thebibliography}

%\begin{thebibliography}{99}
%\end{thebibliography}

%%%%%%%%%%%%%%%%%%%%%%%%%%%%%%%%%%%%%%%%%%%%%%%%%%

%%%%%%%%%%%%%%%%% APPENDICES %%%%%%%%%%%%%%%%%%%%%
\if0
\appendix

\section{Some extra material}

If you want to present additional material which would interrupt the flow of the main paper,
it can be placed in an Appendix which appears after the list of references.
\fi
%%%%%%%%%%%%%%%%%%%%%%%%%%%%%%%%%%%%%%%%%%%%%%%%%%

% Don't change these lines
\bsp	% typesetting comment
\label{lastpage}
\end{document}